\documentclass[twocolumn,amsmath,aps,fleqn]{revtex4}

\usepackage{amssymb}
\usepackage{amsmath}
\usepackage{epsfig}
\usepackage{subfigure}
\usepackage{mathrsfs}
\usepackage{longtable}

\newcommand{\be}{\begin{eqnarray}}
\newcommand{\ee}{\end{eqnarray}}
\newcommand{\bea}{\begin{eqnarray}}
\newcommand{\eea}{\end{eqnarray}}

\newcommand{\adotoa}{\ensuremath{{\cal H}}}

\begin{document}

\title{Phi Zeta Delta: Growth of Perturbations in Parameterized Gravity \newline for an Einstein-de Sitter Universe}
\author{Tessa Baker}
\email{tessa.baker@astro.ox.ac.uk}
\affiliation{Astrophysics, University of Oxford, DWB, Keble Road, Oxford, OX1 3RH, UK}


\begin{abstract}
Parameterized frameworks for modified gravity are potentially useful tools for model-independent tests of General Relativity on cosmological scales. The toy model of an Einstein-de Sitter (EdS) universe provides a safe testbed in which to improve our understanding of their behaviour. We implement a mathematically consistent parameterization at the level of the field equations, and use this to calculate the evolution of perturbations in an EdS scenario. Our parameterization explicitly allows for new scalar degrees of freedom, and we compare this to theories in which the only degrees of freedom come from the metric and ordinary matter. The impact on the Integrated Sachs-Wolfe effect and canonically-conserved superhorizon perturbations is considered.

\end{abstract}

\maketitle
\section{Introduction}
\label{section:intro}

A successful theory of modified gravity has proved hard to find. Most theories begin by postulating new physical principles, for example, the existence of fundamental fields with certain symmetry properties \cite{Nicolis_Galileons} or additional dimensions \cite{Rubakov, MaartensKoyama}; see \cite{MG_report, Nojiri_Odintsov_review} for comprehensive surveys of the current literature. By introducing such principles on which to base a theory we are immediately selecting specific directions in theory space to investigate. An alternative approach is to cautiously explore outwards from the corner of theory space that we understand best, that is, General Relativity (GR). One way to implement this strategy is to construct a parameterized framework that allows for small deviations from GR in a cosmological context, akin to the Parameterized Post-Newtonian framework that has been used to test GR extensively within the Solar System \cite{Will1971, Thorne_Will, Will_Nordvedt_1972, Will2011}.

In order to extract maximum benefit from this approach we need to develop a sense of how the parameters we use impact the growth of structure. 
Historically, our theories of structure formation have been developed using matter-dominated cosmological models \cite{Landau, Peebles, Peebles_book, Blumenthal, Meszaros}. Though not appropriate for realistic calculations, a toy Einstein-de Sitter (EdS) cosmology remains an immensely useful testing ground for new theories; here, we can gain a qualitative understanding of perturbation evolution whilst our knowledge of the background expansion remains on a firm footing. Therefore, maintaining the principle of caution advertised above, we will consider the effects of parameterized gravity on an EdS universe (which contains pressureless matter only). This will be an excellent approximation to the matter epoch of the real universe, and has the added advantage that analytic solutions are achievable in some cases due to the simple properties of pressureless matter.

Despite the wide variety of modified gravity theories present in the literature, a survey of field equations reveals some common features. In many theories the gravitational constant appearing in the Poisson equation acquires a time-dependence and/or scale-dependence \cite{WiggleZ_Nesseris}, and the Newtonian gravitational potential $\Psi$ and curvature perturbation $\Phi$ (in the conformal Newtonian gauge) are not equal as they are in GR. Parameterized frameworks for modified gravity are usually constructed to incorporate such properties. We can ask what typical effects these generic features might have on observables: for example, in addition to affecting the growth rate of structure, distortion of gravitational potentials will imprint secondary anisotropies on the CMB. However, we will assume that modifications to GR must be negligible at very early times, to avoid significantly impacting the primary CMB and the sensitive reaction rates of Big Bang Nucleosynthesis \cite{Bambi_Giannotti,Calabrese,SPT_EDE}; see \cite{Brax_Davis} for consideration of scalar-field models that are non-negligible at the time of recombination. 

This paper investigates the evolution of cosmological perturbations in the parameterized framework implemented in \cite{Skordis2009, Ferreira_Skordis, Baker2011, Zuntz2011}, for an EdS background. 
\textsection\ref{section:param_choice} introduces the necessary formalism in the context of  theories that are constructed purely from metric quantities. In \textsection\ref{section:delta} we use this framework to calculate the evolution of density perturbations on intermediate and large scales. Two other quantities of interest are also calculated: the growth function $f(z)$, and the Integrated Sachs-Wolfe effect that is induced in such gravity theories (note that this is not the same as the late-time Integrated Sachs-Wolfe effect that occurs in a $\Lambda$-dominated era). 

In \textsection\ref{section:extra_dof} we take the first steps towards implementing a similar treatment of gravitational theories that introduce additional scalar degrees of freedom. New degrees of freedom cause a significant increase in complexity which renders a general, model-independent calculation almost impossible, at least analytically. We will implement a parameterized effective fluid approach known as `Generalized Dark Matter' (GDM) \cite{HU_GDM} to facilitate the treatment of these additional scalars, and consider the case in which the effective fluid has a negligible equation of state. This approximation excludes some classes of theories from our analysis, such as $f(R)$ gravity, but is applicable in other cases; we drop the restriction again in \textsection\ref{section:zeta} . In \textsection\ref{section:zeta} we consider a perturbation that is conserved on super-horizon scales in GR, and ask under what conditions this fact remains true in parameterized modified gravity. The conclusions of this work are presented in \textsection\ref{section:discussion}.

The purpose of this paper is to develop an understanding of parameterized gravity, not to pursue accurate calculations for the real universe. 
Hence the plots and trial parameter values used here are intended to be illustrative rather than realistic.

\section{Parameterization of Metric-Only Gravity Theories} 
\label{section:param_choice}
No standard parameterization of modified gravity currently exists. A common choice is to introduce a free function that describes any time- or scale-variation of the gravitational constant in the Poisson equation \cite{Bean, DanielLinder,Pogosian_parameterization,Hojjati_Zhao_2011,Bertschinger_Zukin,Hu_Sawicki}:
\begin{equation}
-2 k^2\Phi=\kappa a^2 \mu (a,k)\,\rho\Delta\label{Poisson}
\end{equation}
where $\kappa=8\pi G_0$, $G_0$ is the canonical value of Newton's gravitational constant, \mbox{$\Delta=\delta+3{\cal H}\theta$} is a comoving gauge-invariant density perturbation and $\theta$ is the velocity potential of a fluid defined by \mbox{$v_i=\nabla_i\theta$}.  
Often a second free function is used to describe the ratio of the two conformal Newtonian potentials, $\eta (a,k)=\Phi/\Psi$. By redefining \mbox{$\zeta=(1-1/\eta)$} this can be rewritten:
\begin{equation}
\Phi-\Psi=\zeta (a,k)\Phi\label{ph_slip}
\end{equation}
The arguments of the `modified gravity functions' (MGFs) $\mu$ and $\zeta$ will be suppressed hereafter. We will call this form of the slip relation parameterization B. In \cite{Baker2011} it was argued that a parameterization of this type is only applicable in the quasistatic regime, and that in the case of purely metric theories it implicitly corresponds to higher-derivative theories. This is true if one assumes that eqn.(\ref{ph_slip}) is an exact `template' for the slip relation of a modified gravity theory, which means that $\zeta$ can only be a function of homogeneous background quantities. If instead one is prepared to let $\zeta$ be a function of non-homogeneous environmental variables and initial conditions then the slip relation may not necessarily imply a higher-derivative theory; it becomes impossible to ascertain the derivative order of the theories being parameterized without further information \cite{private_comm}.

In \cite{Baker2011} an alternative format was suggested, in which the order of derivatives in the field equations is made explicit. In this alternative parameterization a metric theory with a $\Lambda$CDM background incurs extra constraint equations (see Table \ref{tab:2nd_order_constraints} in the appendix), which force the slip relation to be \cite{Skordis2009, Ferreira_Skordis}:  
\begin{equation} 
\Phi-\Psi=\zeta\Phi+\frac{(\mu-1)}{{\cal H}\mu}\dot\Phi
\label{sf_slip}
\end{equation}
This slip relation will be termed parameterization A. The key feature to note here is that in parameterization A the MGF $\mu$ appears in both the Poisson and slip equations, unlike parameterization B. This apparently small difference leads to some degree of ambiguity in the interpretation of current constraints on the MGFs \cite{Zuntz2011}.

However, the purpose of this paper is not to discuss the subtleties of parameterization choice at length; we wish to keep our results as general as possible. We can treat both of the parameterizations simultaneously by adopting the following Poisson and slip equations:
\begin{eqnarray}
-2 k^2\Phi&=&\kappa a^2 \mu_P \,\rho_M\Delta_M\label{gen_Poisson}\\
\Phi-\Psi&=&\zeta\Phi+\frac{\mu_s-1}{{\cal H}\mu_s}\dot\Phi\label{gen_slip}
\end{eqnarray}
To recover parameterization B we set $\mu_s=1$ but keep $\mu_P$ general. To recover parameterization A we set $\mu_s=\mu_P$. Note that $\mu_s$ is related to the function $\tilde{g}$ in \cite{Baker2011} by \mbox{$\mu_s=(1-\tilde{g})^{-1}$}. Throughout this section we will sometimes leave results expressed in terms of the three functions $\mu_P,\,\mu_s$ and $\zeta$. We wish to emphasize from the outset that there are only really two independent functions, and all expressions should be evaluated in either the A-type or B-type instance. We write our expressions in this general format because can be instructive to see whether the modified terms have their origin in the Poisson equation (indicated by the presence of $\mu_P$) or the slip relation (indicated by $\mu_s$ and $\zeta$).

What theories map onto equations (\ref{gen_Poisson}) and (\ref{gen_slip})? The answer is `very few', which is a cause for concern given that the above forms are often used to obtain constraints on modified gravity from current data. To map \textit{exactly} onto these parameterizations a theory must stem from an action that is constructed only from curvature invariants and leads to fields equations that contain at most second- or third-order time derivatives (for parameterizations A and B respectively). By the immense power of Lovelock's theorem \cite{Lovelock1, Lovelock2} such a theory can only differ from GR if it introduces either nonlocality or spacetimes of dimension greater than four. This is a very restricted class of theories, though examples do exist \cite{Deser_Woodard, Deffayet_Woodard, Zumino}.

The limitations described above can be relaxed if one is satisfied with an approximate correspondence between theories and parameterization, rather than an exact one. For example, it is frequently assumed that the perturbed Einstein equations will retain the form of eqns.(\ref{Poisson}) and (\ref{ph_slip}) in theories with additional scalar degrees of freedom. This cannot be exactly true; any new scalar coupled to gravity will modify the zeroth-order Einstein equations in some way, and we expect to see perturbations of the new field appearing in the linearized Einstein equations. However, the form of eqns.(\ref{Poisson}) and (\ref{ph_slip}) is retained within a limited range of distance scales for some theories \cite{Baker2011,Amendola_WL,Schimd,Pogosian_Silvestri,KoyamaMaartens_structureDGP}.

To avoid such approximations we will proceed by taking eqns.(\ref{gen_Poisson}) and (\ref{gen_slip}) at `face value', i.e. assuming that there are no new scalar degrees of freedom hidden behind them. This is the assumption that is implicitly being made if equations such as (\ref{Poisson}) and (\ref{ph_slip}) are implemented in an Einstein-Boltzmann solver \cite{MGCAMB,Hojjati_Pogosian,Hojjati_Zhao_2011,Dossett_2011} and used to generate ISW and matter power spectra. 
In \textsection\ref{section:extra_dof} we will introduce an extended parameterization that attempts to account for the additional scalars explicitly.

\section{Density Perturbations}
\label{section:delta}

In this section we will consider how the growth of cold dark matter (CDM) density perturbations in an EdS universe is influenced by the MGFs. This is a model for the matter-dominated epoch of the real universe. The growth of structure during an epoch in which dark energy is also relevant was investigated in \cite{Ferreira_Skordis}, using a specific ansatz for $\mu_P$ and $\zeta$.

We will assume that any time-variation of the MGFs during the matter-dominated epoch must be very small in order to prevent them from evolving to a region of parameter space that would cause conflict with observations. This is not guaranteed, for example, in \cite{Zuntz2011} we found that cancellations between MGFs can lead to observables very close to the predictions of $\Lambda$CDM. Hence even very radical departures from GR can be accommodated by current data in finely-tuned situations; however, for the purposes of this paper we will assume that our present universe does not correspond to such a case. We will therefore take the time derivatives of  $\mu_P, \mu_s$ and $\zeta$ to be negligible in comparison to the rate of evolution of other variables. We intend to relax this restriction in future work.

Although we are neglecting their time-dependence, $\mu_P$ and $\zeta$ may still contain scale-dependence. They are dimensionless functions, but scale-dependence can appear as a ratio to some special scale that arises in a given theory, i.e. \mbox{$k/k_*$}. An example of such a privileged scale is the Compton wavelength of the scalaron in $f(R)$ gravity \cite{Sotiriou_fR}.

The fluid conservation equations for CDM energy density and momentum perturbations are given by:
\begin{eqnarray}
\dot\delta_M&=&-k^2\theta_M+3\dot\Phi\label{CDM_delta}\label{fluid_eqn1}\\
\dot\theta_M&=&-{\cal H}\theta_M+\Psi\label{CDM_theta}\label{fluid_eqn2}
\end{eqnarray}
where $\theta_M$ is the velocity potential. Differentiating eqn.(\ref{CDM_delta}) and combining with eqn.(\ref{CDM_theta}) leads to a second-order equation for $\delta_M$:
\begin{equation}
\label{delta_evol_1}
\ddot\delta_M+{\cal H}\dot\delta_M - 3\ddot\Phi-3{\cal H}\dot\Phi+k^2\Psi=0
\end{equation}
This is the same as found in GR. However, differences from GR arise when we use a non-trivial slip relation to eliminate $\Psi$. 
Substituting eqn.(\ref{gen_slip}) into eqn.(\ref{delta_evol_1}):
\begin{equation}
\hspace{-0.4cm}
\label{delta_evol_2} 
\ddot\delta_M+{\cal H}\dot\delta_M - 3\ddot\Phi-3{\cal H}\dot\Phi\left[1+\frac{k^2}{3{\cal H}^2}\left(\frac{\mu_s-1}{\mu_s}\right)\right]+k^2(1-\zeta)\Phi=0
\end{equation}
Before studying the behaviour of this equation it is useful to delineate a hierarchy of distance scales:
\begin{enumerate}
\item The nonlinear scale on which clusters and galaxies form.
\item The quasistatic scale on which time derivatives of perturbations can be neglected in comparison to their spatial derivatives.
\item Larger scales on which the above approximation is no longer valid, but are still well within the horizon.
\item Scales that are greater than our observable horizon. 
\end{enumerate}
We will consider the solutions of eqn.(\ref{delta_evol_2}) in regions 3 and 4. In fact it is possible to derive a single equation for $\Phi$ that is valid in both regions 3 and 4, and then use the Poisson equation to relate its solutions to $\delta_M$ (we thank C. Skordis for pointing this out). We will use an equivalent method that is simpler but a little less elegant.

\subsection{Subhorizon scales}
\label{sub:subhorizon_scales}
In region 3 described above we can approximate \mbox{${\cal H}/k  \ll 1$} and \mbox{$\Delta_M\approx\delta_M$} since \mbox{$|\theta_M|\sim |v_M|/|k|$} is small. We use derivatives of eqn.(\ref{gen_Poisson}) to eliminate $\Phi$ from eqn.(\ref{delta_evol_2}): 
\begin{eqnarray}  
&&\ddot{\delta}_M+{\cal H}\dot\delta_M\left[1+\frac{3}{2}\mu_P\frac{(\mu_s-1)}{\mu_s}\right]\\
&&\hspace{2.0cm}-\frac{3}{2}{\cal H}^2\mu_P\delta_M\left[1-\zeta+\frac{(\mu_s-1)}{\mu_s}\right]=0 \nonumber
\label{delta_evol_3}
\end{eqnarray}
where we have used the result $\dot\adotoa=-\frac{1}{2}\adotoa^2$ in an EdS universe. Writing the solutions of this equation in the form
 \begin{equation}
 \delta_M = N^{+} \,a^{\frac{n^{+}}{2}}+N^{-}\,a^{\frac{n^{-}}{2}}
\end{equation}
where $N^{+}$ and $N^{-}$ are constants and the power-law indices are:
\begin{eqnarray}
\label{n_general}
n^{\pm}&=&-\frac{1}{2}\left(1+3\mu_P\frac{(\mu_s-1)}{\mu_s}\right)\\
&&\pm\frac{1}{2}\left[9\frac{\mu_P^2}{\mu_s^2}(\mu_s-1)^2-30\frac{\mu_P}{\mu_s}+6\mu_P(9-4\zeta)+1\right]^{\frac{1}{2}} \nonumber
\end{eqnarray}
In parameterization A this reduces to
\begin{equation}
n^{\pm}_{A}=\left(1-\frac{3}{2}\mu\right)\pm\frac{1}{2}\sqrt{9\mu^2+12\mu(3-2\zeta)-20}\label{n_SF}
\end{equation}
whereas in parameterization B it becomes
\begin{equation}
n^{\pm}_{B}=-\frac{1}{2}\pm\frac{1}{2}\sqrt{1+24\mu_P(1-\zeta)}\label{n_phenom}
\end{equation}
It can be verified that in the limit $\mu_s=\mu_P=1$ and $\zeta=0$ eqn.(\ref{n_general}) recovers the GR result $\delta_M \propto a$. We can see immediately that in both parameterizations there is a term $-24\mu_P\zeta$ that leads to degeneracy between the effects of the individual MGFs. Note that $\zeta$ only appears within this degenerate combination, so it cannot significantly impact growth if $\mu_P$ is small. We note that if a $\mu$-like MGF is implemented in the Poisson equation for $\Psi$ instead of $\Phi$ then this degeneracy does not arise in the parameterization B case \cite{Pogosian_Silvestri, Hojjati_Pogosian,Hojjati_Zhao_2011}.

When $n^{\pm}$ are imaginary the solutions for $\delta_M$ are damped oscillations. However, since our calculation has neglected the effects of baryons or radiation this oscillatory behaviour simply indicates unphysical solutions rather than anything meaningful. To have at least one growing mode we need $n^{+}$ to be positive, for which the relevant condition is:
\begin{equation}
\label{growing_bound}
\mu_s (2-\zeta) > 1
\end{equation}
We expect that an approximate version of this bound should be obeyed in the real universe, in order to reproduce the observed matter power spectrum - see \textsection\ref{sub:constraints}. However, in the real universe the hard bound of eqn.(\ref{growing_bound}) will be softened by contributions to growth from the radiation and $\Lambda$-dominated eras. Note that there is no restriction that prevents $\zeta$ from adopting negative values.

We wish to understand the physical mechanisms through which the MGFs are exerting their influence on small scales. We can get a feel for this by thinking about eqn.(\ref{delta_evol_3}) in the context of a simple mechanical system. The last term on the righthand side represents a time-dependent forcing that drives the collapse of density perturbations. The $\dot{\delta}_M$ term is analogous to a frictional force, which in familiar physical situations always acts to oppose motion; its magnitude decreases with time due to the factor of $\adotoa$. Since we are parameterizing around a $\Lambda$CDM background the evolution of $\adotoa$ is unaffected by the MGFs, and hence cannot be contributing to deviations from GR. 

The overall magnitude of the driving term is controlled by $\mu_P$. This intuitively makes sense -- if we increase the gravitational coupling strength then we expect structures to collapse faster. Less intuitive is the appearance of $\mu_s$ and $\zeta$ in the driving term, which have the ability to change its sign. We will assume throughout that \mbox{$\mu_P,\, \mu_s>0$} always to maintain agreement with our physical notion of attractive gravity, but note that there is no such restriction on $\zeta$. The condition for the driving force to maintain the same direction as it has in GR is exactly eqn.(\ref{growing_bound}). It is interesting to see that in parameterization A $\mu_s$ and $\zeta$ can have counteracting effects on the driving term. 
Qualitatively, a negative $\zeta$-value with large magnitude enables one to weaken $\mu_P$ considerably whilst maintaining growth during a matter-dominated epoch. If parameterization B is adopted this effect does not exist because the modification to the Poisson equation has no influence on the sign of the driving term.

The friction term has a somewhat simpler behaviour, as it is unaffected by $\zeta$. In parameterization B the friction is unchanged from GR, but in parameterization A \mbox{$\mu=\mu_P=\mu_s$} has the power to enhance or suppress friction effects. A value of $\mu>1$ acts to increase friction, but simultaneously strengthens the driving term that drives perturbations to collapse. One expects that some degree of cancellation between these two effects may be possible, even without the extra freedom provided by $\zeta$.

\subsection{Superhorizon scales}
\label{sub:superhorizon_scales}
On large scales we ought not to make the approximation \mbox{$\Delta_M\approx \delta_M$}, as the magnitude of the velocity potential is not negligible. We will adopt a different strategy, using the linearized Friedmann equation to solve for $\Phi$ as a proxy for $\delta_M$. 

We first adopt a phenomenological approach (in the spirit of parameterization B), and assume that Newton's constant is identically modified in all perturbed field equations. We then have (neglecting a small term proportional to $k^2$) :  
 \begin{equation}
-{\cal H}(\dot\Phi+{\cal H}\Psi)=\frac{\kappa a^2}{6} \rho_M\mu_P\delta_M=\frac{{\cal H}^2}{2}\mu_P\delta_M\\
\end{equation}
Differentiating, and using that $\dot\delta_M\approx 3\dot\Phi$ on large scales (from eqn.(\ref{fluid_eqn1})):
\begin{equation}
\ddot{\Phi}+\frac{\cal H}{2}(1+3\mu_P)\dot\Phi+{\cal H}\dot\Psi=0
\end{equation}
Using the slip relation (eqn.(\ref{gen_slip})):
\begin{equation}
\ddot\Phi+\dot\Phi\frac{\cal H}{2}\left[3+3\mu_P-2\zeta\right]=0
\end{equation}
The power-law solutions are $\Phi\propto a^{p^{\pm}}$, with $p^{+}=0$ and \mbox{$p^{-}=(2\zeta-2-3\mu_P)$}. On these scales $\delta_M$ follows the behaviour of $\Phi$ up to a constant offset, which can be set to zero by initial conditions. So the dominant mode -- constant potential outside the horizon and $\delta_M$ frozen -- is the same as in GR. However, the decaying mode is affected by the MGFs. For $\zeta=0$, increasing $\mu_P$ will result in faster decay. This seems somewhat counterintuitive, since one would usually associate an increase in gravitational strength with \textit{reduced} decay of overdensities. However, since we are working on superhorizon scales gauge issues may invalidate our physical notions of gravitational growth and decay.

In parameterization A the linearized Friedmann equation is
 \begin{equation}
-6{\cal H}(\dot\Phi+{\cal H}\Psi)=\kappa a^2 \rho_M\delta_M+A_0 k^2\Phi \label{00E_SF}
\end{equation}
where
\begin{eqnarray}
A_0&=&-2\left(\frac{\mu_P-1}{\mu_P}\right)\left(1+\frac{{\cal H}^2}{Q}\right) +2\zeta\frac{{\cal H}^2}{Q}\nonumber\\  
\mathrm{with}\quad\;Q&=&{\cal H}^2+\frac{k^2}{3}-\dot{\cal H}\nonumber
\end{eqnarray}
The derivation of the above expression is described in appendix \ref{app:Z_consv}.

Note that $\mu_P$ does not feature explicitly in eqn.(\ref{00E_SF}). Newton's constant is not modified directly -- instead one considers all possible additional terms that could appear in the linearized Einstein equations, which can be determined up to a dimensionless function of background quantities . In the case of second-order metric-only theory the only terms that can be added to the linearized Friedmann and `0i' equations are proportional to $\Phi$, which can be absorbed into an effective Newton's constant.

On very large scales $\lim_{k\to 0}(k^2A_0)=0$, so we can neglect the extra term on the RHS of eq.(\ref{00E_SF}). Repeating the steps we took for the parameterization B-like case we obtain $p^+=0, \,p^-=\mu_P\left(2\zeta-5\right)$. The values of $p^{-}$ in the two parameterizations converge as one tends to the GR limit, as of course they must.

However, the decaying mode is not hugely interesting as it is unobservable (unless there are some very radical modifications to GR involved). The important result here is that the potential remains constant on superhorizon scales, as usual.

\subsection{Connection to Constraints}
\label{sub:constraints}
\begin{figure}[t!]
\begin{center}
\includegraphics[width=9.2cm]{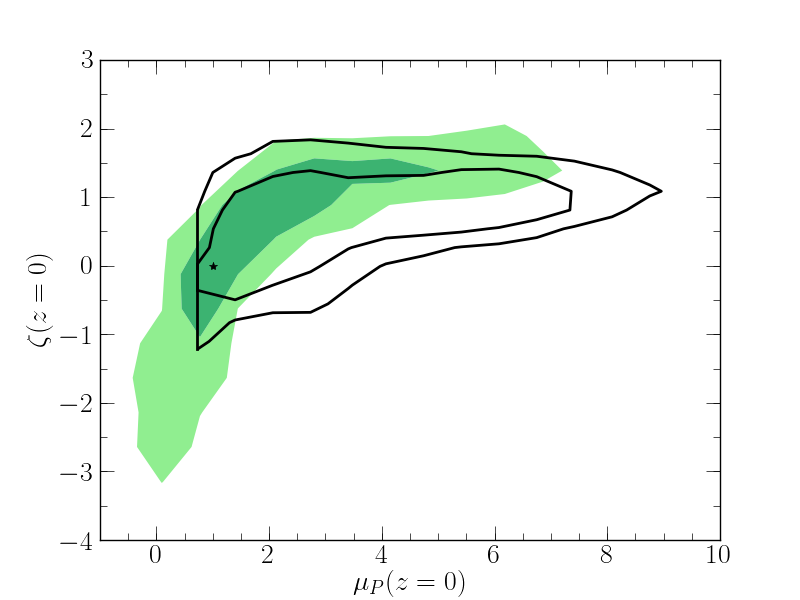}
\caption{Joint constraints on the slip parameter $\zeta$ and $\mu_P$  at $z=0$, for parameterization A (black lines) and B (filled green).  In both cases 68\% and 95\% contours are shown.}  
\label{constraints_plot}
\end{center}
\end{figure}

To what extent are our results for an idealized EdS model borne out in the real universe? Fig.\ref{constraints_plot} shows the joint constraints on the MGFs $\mu_P$ and $\zeta$ obtained using the following data sets: the 7-year WMAP CMB data \cite{wmap7}, the SDSS DR7 matter power spectrum \cite{sdss-dr7}, a prior $H_0 = 73.8 \pm 2.4$ \cite{riess2011}, the BBN constraint $\Omega_b h^2 = 0.022 \pm 0.002$ \cite{bbn}, and the Union2 Supernova Ia data \cite{Amanullah2010}. The differences between the two sets of contours were discussed in detail in \cite{Zuntz2011}. Here our main interest is the extent to which the constraints reflect the analytic solutions of the previous two subsections. 

Given the restriction that $\mu_s$ must be positive, eqn.(\ref{growing_bound}) tells us to expect that $\zeta < 2$ on subhorizon scales in parameterization A. For $\zeta=0$ we must have $\mu_P=\mu_s>1/2$ for density perturbations to grow during the matter era, which corresponds to the approximate location of the near-vertical black contour. This contour is not the result of any artificially-imposed boundary, but delineates a very sharp fall-off in the likelihood distribution for $\mu_P$ (see figure 6 of \cite{Zuntz2011}). In contrast, in parameterization B any $\mu_P>0$ permits growing modes for $\zeta=0$, giving rise to the more gradual fall-off shown by the shaded countours.

However, eqn.(\ref{growing_bound}) also implies that in parameterization B $\zeta<1$ is necessary for growth of CDM density perturbations on  subhorizon scales, which is contradicted by Fig.\ref{constraints_plot}. This is not too surprising -- we expect the simple bounds implied by our EdS example to be blurred by the complexities of a realistic cosmological model. If models in the region of parameter space $\zeta>1$ experience sufficient growth during radiation- and $\Lambda$-dominated epochs, or on scales outside the validity of eqn.(\ref{n_phenom}), then they will not be excluded by an MCMC analysis.

The degeneracy between $\mu_P$ and $\zeta$ is visible in both contour sets, but it is more pronounced in parameterization B. This is because in parameterization A the quadratic term in eqn.(\ref{n_SF}) makes it more difficult to accommodate the effects of a large $\mu_P$ with a small $\zeta$, or vice-versa.

\subsection{Other Growth Observables}
\label{sub:growth_observables}

\subsubsection*{Integrated Sachs-Wolfe Effect}
\label{subsub:ISW}

A cosmological model governed by GR will not experience an Integrated Sachs-Wolfe (ISW) effect \cite{Sachs_Wolfe} during a matter-dominated epoch. The kernel of interest for the ISW effect is $\dot\Phi+\dot\Psi$ (in the conformal Newtonian gauge), which in GR is equal to $2\dot\Phi$. From the standard Poisson equation one has, on scales well below the horizon:
\begin{equation}
\Phi\propto a^2\rho_M\delta_M
\end{equation}
Energy-momentum conservation gives $\rho_M\propto a^{-3}$, whilst in GR $\delta_M\propto a$, leaving $\Phi$ with zero time-dependence. The situation only changes for $z \lesssim 0.5$ when $\Lambda$ begins to suppress the growth rate of $\delta_M$ \cite{Crittenden_Turok}. 

In modified gravity this behaviour is affected in two ways: a non-trivial slip relation will cause the ISW kernel to differ from $2\dot\Phi$, and the scaling of $\delta_M$ with $a$ will be altered. Then generically we expect a non-zero ISW effect, even during a matter-dominated phase of the universe \cite{Zhang_2006, Giannantonio_2010}. We will calculate the contribution to the CMB temperature power spectrum of this effect for the metric theories discussed in \textsection\ref{section:delta}.

We begin by using the slip relation (eqn.(\ref{gen_slip})) to express the ISW kernel purely in terms of $\Phi$:
\begin{equation}
\label{ISW1}
\dot\Phi+\dot\Psi=\frac{1}{2}\dot\Phi\left[3-2\zeta+\frac{1}{\mu_s}\right]-\ddot\Phi\left(\frac{\mu_s-1}{{\cal H}\mu_s}\right)
\end{equation}
Using the Poisson equation to connect $\Phi$ and $\delta_M$, we have (discarding the decaying mode):
\begin{eqnarray}
\Phi&=&M(k)\eta^{(n^{+}-2)}\nonumber\\
\mathrm{where}\;\;M(k)&=&-\frac{\kappa \delta_{M,0}(k)\mu_P \rho_{M,0}}{2k^2}\nonumber
\end{eqnarray}
$n^{+}$ is given by eqn.(\ref{n_general}) and density perturbations are normalised by their present values i.e. \mbox{$\delta_M (k,\eta)=\delta_{M,0}(k) \,\eta^{n^{+}}$}. Generally one expects the growth rate of density perturbations to be scale-dependent in a modified gravity theory, whereas in GR all linear subhorizon modes grow at the same rate. However, we are modelling a region of theory space close to GR and hence we will assume negligible variation of the growth rate over the range of $k$ relevant to observations of the ISW plateau.

So the ISW kernel is:
\begin{eqnarray}
\dot\Phi+\dot\Psi&=&\frac{M(k)}{2}(n^{+}-2)\eta^{(n^{+}-3)}\times\nonumber\\
&&\left[6-n^{+}-2\zeta+\frac{1}{\mu_s}(n^{+}-2)\right]\label{ISW2}
\end{eqnarray}
Note that in the GR limit $n^{+}=2$ the above expression vanishes as expected. 

Next we need to compute the power spectrum of this modified-gravity induced ISW effect. The expression to be evaluated is:
\begin{equation}
C_l=\frac{2}{\pi}\int_0^{\infty}dk\,k^2P(k)\,\Big| \frac{\Theta_l(k)}{\delta_{M,0} (k)}\Big|^2
\label{general_Cl}
\end{equation}
The temperature perturbation observed today, $\Theta_l(k, \eta_0)$, consists of three parts:
\begin{eqnarray}
\Theta (k,\eta_0)&=&\:\mathrm{monopole\;term}\:+\:\mathrm{dipole\;term}\:\\
&&\hspace{-0.5cm}+\int^{\eta_0}_0d\eta\:e^{-\tau}\left[\dot\Phi (k,\eta)+\dot\Psi(k,\eta)\right] j_l[k(\eta_0-\eta)] \nonumber
\end{eqnarray}
Let us focus on the dominant contribution to the ISW power spectrum which comes from the  $(\dot\Phi+\dot\Psi)^2$ term, and ignore the cross terms with the monopole and dipole. The cross-terms should only yield small corrections because the monopole and dipole terms are evaluated at the time of last scattering, and hence affect different $l$-values from the subsequent ISW effect. We will take the visibility function $e^{-\tau}$ to be a step function at recombination. Of course, if we are considering a truly EdS universe then there is no recombination event or time of last scattering, but this detail is irrelevant -- we are only interested in modelling the real universe well after recombination. In the real universe recombination occurs sufficiently early that for practical purposes we can take $\eta_{\mathrm{rec}}\approx 0$. 

A standard derivation relates the power spectrum of density fluctuations today to the power spectrum of the primordial potential \cite{Dodelson}. The only modification to this that occurs in our theory is a factor of $\mu_P^{-2}$, which arises when we relate density fluctuations to $\Phi$ via the Poisson equation. However, this is cancelled by a factor of $\mu_P^2$ in $M(k)$. The scales of interest to us are sufficiently large that we can set the transfer function $T(k)\sim 1$. Then, for a Harrison-Zel'dovich spectrum we find
\begin{eqnarray}
\label{Cl_ISW}
\hspace{-0.2cm}C_l^{ISW,sq}&=&\frac{9\pi}{4}(n^{+}-2)^2 \left(6-n^{+}-2\zeta+\frac{1}{\mu_s}(n^{+}-2)\right)^2\delta_H^2\nonumber\\
&&\hspace{-5mm}\times\int_0^{\infty}\frac{dk}{k}\left[\int^{\eta_0}_0\,\eta^{(n^{+}-3)}j_l[k(\eta_0-\eta)]\,d\eta\right]^2
\end{eqnarray}
where $\delta_H$ is the amplitude of primordial perturbations at horizon-crossing during a single-field slow-roll inflation scenario. 
The superscript `$sq$'  on $C_l$ reminds us that we have only evaluated the ISW-squared term and not the cross-terms. 

\begin{figure}
\includegraphics[scale=0.5]{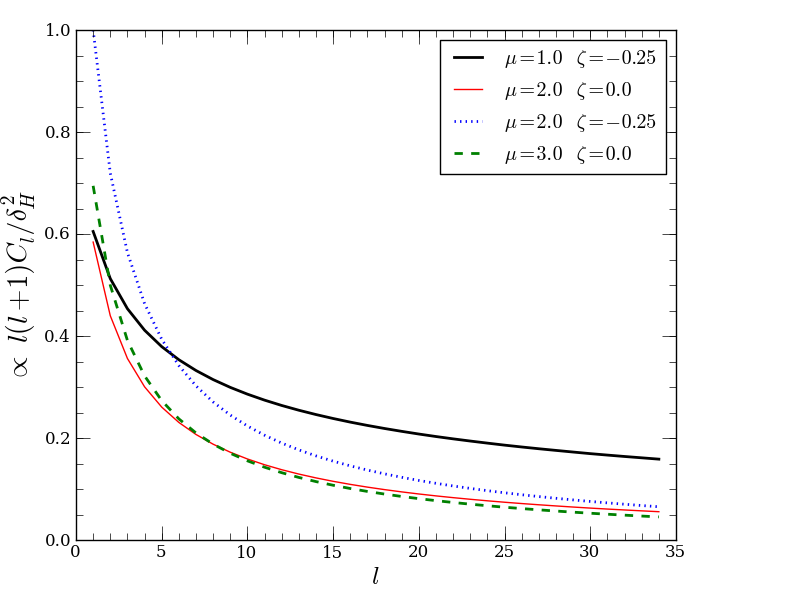}
\caption{Low-$l$ power spectrum of the ISW effect induced by a theory of modified gravity constructed in parameterization A (see \textsection\ref{section:param_choice} for details). The y-axis scaling is arbitrary. The region $l\lesssim 10$ is expected to be subject to corrections.}
\label{fig:ISW}
\end{figure}

A plot of this power spectrum for several combinations of $\mu$ and $\zeta$ in parameterization A is shown in Fig.\ref{fig:ISW}. The normalization of the y-axis is arbitrary because we have not attempted an accurate calculation of the $C_l$s; we are more interested in how the shape and amplitude of the power spectrum is affected by different parameters. Note that our subhorizon solution for $\Phi$ is likely to become invalid for the largest scales, hence the region $l\lesssim 10$ in Fig.\ref{fig:ISW} is not fully accurate.

Provided that $\mu_P$ is not unusually small, $\zeta$ can have a significant effect on the amplitude of the power spectrum. For example, compare the curves with $\{\mu_P,\zeta\}$ equal to $\{1,-0.25\}$ and $\{2,0\}$ -- a small change in $\zeta$ dominates over a large change in $\mu_P$. 
The models $\{2.0,-0.25\}$ and $\{3.0,0.0\}$ have the same value of $n^+$, and hence the same spectral shape. It may seem a little surprising that a model with parameters $\{1.0,-0.25\}$ predicts a larger ISW effect than one with $\{2.0,-0.25\}$ for $l>6$; this is because a larger value of $n^+$ shifts the dominant contribution to the integral in eqn.(\ref{Cl_ISW}) to later times, therefore shifting the corresponding power spectrum left towards larger scales.

In general, however, it is difficult to cleanly disentangle the effects of $\mu_P$ and $\zeta$ on the ISW power spectrum, because they appear in a degenerate combination inside $n^+$.

\subsubsection*{Growth Function}
\label{subsub:growth_function}
The rate of growth of structure as a function of redshift is often quantified via the \textit{growth function}:
\begin{equation}
\label{growth_function}
f(z)=\frac{d\,\mathrm{ln}\,\Delta_M}{d\,\mathrm{ln}\,a} 
\end{equation}
In GR $f(z)$ is independent of the wavenumber $k$, but in a modified gravity scenario this is not generally the case. However, if we assume we are not dealing with very radical departures from GR then this scale-dependence is likely to be small over a restricted range of $k$. On the scales of region 3 (see the beginning of this section) where $\Delta_M\sim\delta_M$, the growth function is simply \mbox{$f(z)=\frac{n^{+}}{2}$}, where $n^{+}$ is given by eqn.(\ref{n_general}).
\vspace{1mm}

\section{Theories With Additional Degrees of Freedom}
\label{section:extra_dof}

Many gravity theories introduce degrees of freedom (d.o.f.) other than perturbations of the metric -- additional scalar or vector fields, second metrics, or any combination thereof \cite{MG_report}. For model-independent constraints to be genuinely feasible we need to construct a parameterization that is able to accommodate such d.o.f. In this paper we will restrict ourselves to scalar d.o.f. only, which find widespread motivation from particle physics, braneworld models and string theory. 

In \textsection\ref{section:param_choice} the parameterization-A-based approach was to add to each linearized Einstein equation all possible terms that could appear in the context of a metric-based second-order theory, which amounted to terms in $\Phi$ and $\dot\Phi$ multiplied by some function of background quantities (see appendix \ref{app:Z_consv}). When we extend this framework to include extra d.o.f. two new types of terms appear. Firstly we must allow for 
 perturbations of the d.o.f themselves -- for example, in a Brans-Dicke theory \cite{BransDicke} one expects the perturbations of the scalar field $\delta\phi,\,\delta\dot{\phi}$ and $\delta\ddot{\phi}$. Such new scalars are awkward to work with directly without knowing their underlying equations of motion. Therefore we will adopt an alternative approach based on a scheme by Hu \cite{HU_GDM} and treat this first type of additional term as perturbations of an effective fluid. A similar approach was adopted in \cite{Koivisto_Mota, Mota_2007}.

In a fully general case one should allow the effective fluid to have a time-varying equation of state, non-adiabatic perturbations and significant anisotropic stress. This would render our simple EdS model invalid by modifying the evolution of the cosmological background. To proceed with our analytic treatment we will sacrifice some generality by setting the equation of state of the effective fluid to zero. It then contributes to the zeroth-order Friedmann equation and supports density and velocity perturbations, but has negligible pressure and anisotropic stress.

This approximation is not unreasonable; for example, in Ho\v{r}ava-Lifshitz gravity \cite{HL1, HL2} a non-local Hamiltonian constraint gives rise to an integration constant which may be interpreted as a pressureless fluid. The additions to the Einstein equations in linear Einstein-Aether theory also behave as a pressureless fluid during an EdS phase, as can be seen using eqns.(15) and (16) of \cite{Zlosnik_structure_growth}. The effective fluid of Eddington-Born-Infeld gravity can experience a CDM-like phase too \cite{Banados_Ferreira_Skordis}. However, making this restriction does exclude some important classes of theories, such as $f(R)$ gravity, except for special choices of $f(R)$.

The second type of new term that arises when extra d.o.f. are included is the following combination of metric potentials:
\begin{equation}
\label{gamma}
\Gamma=\frac{1}{k}\left(\dot\Phi+\adotoa\Psi\right)
\end{equation}
In this paper we are working in the conformal Newtonian gauge, but the parameterization we are using is constructed using gauge-invariant quantities -- see \cite{Skordis} for details. In the gauge-invariant formalism $\Phi$ and $\Psi$ are replaced by the Bardeen potentials $-\Psi_H$ and $\Phi_A$. If the background equations are fixed to be $\Lambda$CDM, as in \textsection\ref{section:delta}, then the gauge-invariant version of $\Gamma$ cannot appear in the linearized field equations because it introduces higher-order time derivatives of the scale factor. However, when new d.o.f. are added these unwanted time derivatives can be cancelled by perturbation of the new scalar. This somewhat technical point was demonstrated explicitly for scalar-tensor theories in appendix B of \cite{Baker2011}. 

Concretely, then, we write the Einstein equation in the form:
\begin{equation}
\label{einstein}
G_{\mu\nu} \;=\; 8\pi G_0 a^2\, T_{\mu\nu}+a^2 U_{\mu\nu}
\end{equation}
where the tensor $U_{\mu\nu}$ contains all the non-standard terms arising from a theory of modified gravity. At the perturbed level the components of $\delta U_{\mu\nu}$ are then given by (in Fourier space):
 \begin{eqnarray}
 \label{2nd_order_Us}
U_{\Delta} &=& A_0 k^2\Phi+F_0k^2\Gamma+\kappa a^2 \rho_E\delta_E\nonumber\\ 
U_{\Theta}&=& B_0 k\Phi+I_0k\Gamma+\kappa a^2 \rho_E\theta_E\nonumber \\
U_P &=& C_0 k^2\Phi + C_1 k\dot{\Phi}+J_0 k^2\Gamma+J_1 k \dot\Gamma  \nonumber\\
U_{\Sigma} &=&  D_0\Phi+\frac{D_1}{k} \dot{\Phi}+K_0\Gamma+\frac{K_1}{k} \dot\Gamma  \label{U_2nd_order}
\end{eqnarray}
where 
\begin{eqnarray}
\label{U_cmpt_def}
U_{\Delta}&=&-a^2\delta U^0_0, \qquad  \vec{\nabla}_i U_{\Theta}=-a^2 \delta U^0_i \nonumber\\
 U_P&=&a^2\delta U^i_i, \qquad\,\, D_{ij}U_{\Sigma}=a^2(\delta U^i_j-\frac{1}{3}\delta U^k_k\delta^i_j)
\end{eqnarray}
The coefficients $A_0,\dots K_1$ above are functions of background quantities, dependencies which we will suppress to avoid cluttered expressions. 
A subscript $E$ denotes the quantities relating to the effective fluid. The expressions above contain fewer terms than the corresponding ones in \cite{Baker2011}; the additional metric terms present in that paper are those needed to form a gauge-invariant combination with perturbations of the new scalar. Here we have kept those terms folded into the effective fluid so that the correspondence with $\delta U_{\mu\nu}$ in a general gauge is clearer.

 The system to be solved then comprises of the six variables \{$\Phi, \Gamma, \delta_E, \theta_E, \delta_M,\theta_M$\} and six dynamical equations: the two spatial components of the linearized Einstein equation and two fluid conservation equations for each of CDM and the effective fluid (equivalent to Bianchi identities). Note that because $\delta U_{\mu\nu}$ contains additional metric perturbations the conservation equations of the effective fluid will contain some non-standard terms. The full system of equations is displayed in appendix \ref{app:dof_eqns}. 

$\Gamma$ can be eliminated from the two spatial Einstein equations to give a second-order equation in terms of $\Phi$:
\begin{eqnarray}
\label{dof_phi_eqn}
\ddot\Phi&+&\dot\Phi\left[\frac{\dot\alpha}{\alpha}-\frac{\dot{W_1}}{W_1}+k\frac{W_2}{\alpha}+\frac{kZ_1}{\alpha}\right]\\
&&+\frac{k}{\alpha}\Phi\left[\dot{W_2}-\frac{\dot{W}_1W_2}{W_1}+k\frac{W_2 Z_1}{\alpha}-k\frac{W_1Z_2}{\alpha}\right]=0\nonumber
\end{eqnarray}
where $\adotoa_k=\adotoa/k$ and
\begin{eqnarray}
\label{alpha_etc_array}
\alpha&=&\left(D_1-\frac{1}{\adotoa_k}\right)(J_1-6)+K_1\left(9\adotoa_k-C_1+\frac{2}{\adotoa_k}\right)\nonumber\\
W_1&=&K_1\left(3\adotoa_k-J_0-\frac{2}{\adotoa_k}\right)+(J_1-6)\left(K_0+\frac{1}{\adotoa_k}\right)\nonumber\\
W_2&=&(J_1-6)(D_0-1)+K_1(2-C_0)\nonumber\\
Z_1&=&\left(\frac{1}{\adotoa_k}-D_1\right)\left(3\adotoa_k-J_0-\frac{2}{\adotoa_k}\right)\nonumber\\
&&+\left(\frac{1}{\adotoa_k}+K_0\right)\left(9\adotoa_k-C_1+\frac{2}{\adotoa_k}\right)\nonumber\\
Z_2&=&(2-C_0)\left(\frac{1}{\adotoa_k}-D_1\right)+(D_0-1)\left(9\adotoa_k-C_1+\frac{2}{\adotoa_k}\right) \nonumber
\end{eqnarray}
We have assumed that time derivatives of the MGFs are negligible, consistent with our treatment of metric-only theories. If we set the MGFs to zero the third term vanishes and we recover the GR result $\Phi=$ constant. Modifications to the Poisson equation mean that the solution for $\Phi$ is not as easily translated into a solution for $\delta_M$ as it is in GR. One route is to substitute the solution for $\Phi$ into either of eqns.(\ref{Einstein3}) or (\ref{Einstein4}) and solve for $\Gamma$, then use both of these solutions in eqn.(\ref{delta_evol_1}) -- see appendix \ref{app:deltas_in_dof} for an explicit calculation.

We wish to consider eqn.(\ref{dof_phi_eqn}) on superhorizon and subhorizon scales, as we did in \textsection\ref{section:delta}. However, there is some uncertainty involved in taking this limit without knowing specifically the functional forms hiding behind the MGFs. We will assume that any scale-dependence appears relative to some preferred scale of a given theory, i.e. as a function of $(k/k_*)$.

\subsection{Subhorizon Scales}
\label{subsub:dof_subhorizon}

Under the assumptions stated above, and retaining only the dominant terms when expanded in powers of $(k/\adotoa)$, on subhorizon scales eqn.(\ref{dof_phi_eqn}) reduces to:
\begin{equation}
\ddot\Phi+\frac{(2D_1-C_1+2K_0-J_0)}{(6-J_1+2K_1)}k\dot\Phi+\frac{(2D_0-C_0)}{(6-J_1+2K_1)}k^2\Phi=0  \label{dof_phi_ss}
\end{equation}
With the change of variable $x=k\eta$ the above equation can be rewritten (using primes to denote derivatives with respect to $x$):
\begin{eqnarray}
\label{dof_phi_ss_2}
&&\Phi^{\prime\prime}+\frac{\beta}{\gamma}\Phi^{\prime}+\frac{(2D_0-C_0)}{\gamma}\Phi=0\\
&&\mathrm{where}\;\beta=(2D_1-C_1+2K_0-J_0),\;\;\gamma=6-J_1+2K_1.\nonumber
\end{eqnarray}
Without taking specific forms of the MGFs we cannot say whether the coefficients of the second and third terms above are positive or negative, only that sufficiently large modifications to gravity have the power to flip their signs (although the magnitude of such non-GR terms is expected to be small in the domain of validity of this equation). However, one might expect \mbox{$(2D_0-C_0)/\gamma>0$} and \mbox{$\beta/\gamma>0$} so that changes in $\Phi$ damp out (and therefore return to the GR-like situation) rather than grow.

Eqn.(\ref{dof_phi_ss_2}) has the form of a simple mechanical oscillator, and will display the usual phenomena of ringing or over-damping depending on the values of the coefficients. Specifically, its behaviour will depend on the value of \mbox{$\beta^2/\gamma-8D_0+4C_0$}, with negative values of this quantity leading to damped oscillations in $\Phi(x)$ and positive values leading to exponentially growing and decaying solutions.  
Let us apply the single boundary condition that the potential is constant on superhorizon scales; we will see shortly that this is likely to remain true in theories with extra d.o.f. We can then determine the subhorizon solution up to an overall constant:
\begin{eqnarray}
\Phi(x)&\propto&e^{m_+x}-\frac{m_+}{m_-}e^{(m_+-m_-)}e^{m_-x}\\
\mathrm{where}\;\;m_{\pm}&=&-\frac{\beta}{2\gamma}\pm\sqrt{\left(\frac{\beta}{\gamma}\right)^2-\frac{4(2D_0-C_0)}{\gamma}}\nonumber
\label{dof_phi_ss_soln}
\end{eqnarray}
At first sight this exponential solution might seem to be a cause for concern, as we would normally expect the growth of $\Phi$ or $\delta_M$ on small scales to follow power-law behaviour. The unusual solution above has arisen because the modifications in eqns.(\ref{2nd_order_Us}) introduce factors of $k$ that dominate over the usual GR terms on small scales. In the toy model considered here we took the MGFs to be dimensionless functions of order one; without such assumptions it is difficult to make any general statements about the growth of perturbations, because we do not know how to assess the relative importance of two factors such as $J_0$ and $3k/\adotoa$. Of course, if we know the specific functional forms hidden behind the MGFs then such assumptions are not necessary, but then we would not be pursuing a model-independent approach.

A more realistic situation would be to take the MGFs to be much smaller in magnitude than the GR terms. However, our intention here is to assess qualitatively the effects that parameterized systems of modifications to gravity have on the growth of structure - a task which becomes difficult when the parameters are taken to be vanishingly small. We will therefore maintain the simple assumptions described above, remembering that in a more realistic model the effects described here would only be manifest as small distortions of a predominantly GR-controlled universe.

Alternatively, we could turn this problem on its head. One reason theories with additional d.o.f. are difficult to work with is because we lose the ability to derive a hierarchy of constraint equations between the MGFs, as we did in the purely metric case (these arise from the Bianchi identities -- see appendix \ref{app:Z_consv}). Can we use the growth of $\Phi(x)$ to infer some replacement constraints?

To clarify, we wish to restore power-law behaviour for $\Phi(x)$. By expanding eqn.(\ref{dof_phi_eqn}) in powers of $k/\adotoa$ we can find the conditions necessary to remove the dominant terms that are causing the exponential solution of eqn.(\ref{dof_phi_ss_soln}). We find these to be:
\begin{eqnarray}
2D_0-C_0=0,\;\quad2D_1-C_1=0,\;\quad2K_0-J_0=0
\label{constraints}
\end{eqnarray}
If the above conditions are satisfied on small scales then eqn.(\ref{dof_phi_eqn}) reduces to:
\begin{eqnarray}
&&\ddot\Phi+\adotoa\dot\Phi\left[1-D_0+\frac{12}{\gamma}\right]+\frac{\adotoa^2}{2}\Phi \left(1-D_0\right)\left[\frac{6}{\gamma}-1\right]=0\nonumber\\
\end{eqnarray}
The solutions are then power laws in $a$ (or $\eta$), as desired:
\begin{eqnarray}
\label{phi_sol_dof_sub}
\Phi (a)&=&Pa^{\frac{q^+}{2}}+Qa^{\frac{q^-}{2}}\nonumber\\
q^{\pm}&=&D_0-\frac{12}{\gamma}-\frac{1}{2}\nonumber\\
&&\pm\left(D_0^2-3D_0-\frac{12 D_0}{\gamma}+\frac{144}{\gamma^2}+\frac{9}{4}\right)^{\frac{1}{2}}
\end{eqnarray} 
In appendix \ref{app:deltas_in_dof} we convert this solution for $\Phi (a)$ into a solution for $\delta_M$ and $\delta_E$. Here we shall simply state the results:
\begin{eqnarray}
\delta_M (a)&=&P k^2\frac{(D_0-1) }{(q^{+}+2)(q^{+}+3)}a^{\frac{q^{+}}{2}+1}\\
\delta_E (a)&=&\frac{P k^2}{6\Omega_E}\left[\frac{6(1-D_0)(1-\Omega_E)}{(q^{+}+2)(q^{+}+3)}-\frac{1}{\mu_P}\right] a^{\frac{q^{+}}{2}+1}
\end{eqnarray}
In the GR limit $\Omega_E\to 0$, $D_0\to 0$ (which sends $q^{+}\to 0$) we recover that $\delta_M$ scales with $a$.

It is interesting to see that the conditions in eqn.(\ref{constraints}) are satisfied in parameterization A of \textsection\ref{section:param_choice} in the limit $k\to\infty$ (the last condition trivially so). The same is true in parameterization B, where $D_1=0$. 
If one wished to implement this gravitational framework into a realistic cosmological model, without necessarily taking the absolute magnitude of the MGFs to be very small, then these are the constraints that must be satisfied to give a reasonable degree of structure formation. Of course, this does not prevent the model from being ruled out by other observables such as the ISW effect. Having restored power-law growth, the effects of this theory on the ISW effect and growth function are expected to be qualitatively similar to those in \textsection\ref{sub:growth_observables}.

\subsection{Superhorizon scales}
\label{subsub:dof_superhorizon}
On superhorizon scales eqn.(\ref{dof_phi_eqn}) simplifies to:
\begin{eqnarray}
&&\ddot\Phi+\frac{k\dot\Phi}{K_1}\left(K_0-\frac{D_1}{3}\right)\\
&&-\frac{k^2\Phi}{18 K_1}\Big[(J_1-12)(D_0-1)+K_1\left(2-C_0\right)\Big]=0\nonumber
\label{dof_large_scales}
\end{eqnarray}
This can be reduced to an oscillator equation in $k\eta$, in complete analogy to the subhorizon case. However, in the superhorizon limit $k\to\infty$ these oscillations become infinitely slow in $k\eta$; effectively, we have that $\Phi$ is a constant. This matches the GR  and metric-only cases.

\section{Conserved Superhorizon Perturbations?}
\label{section:zeta}

Any relativistic theory of gravity in which energy-momentum is covariantly conserved allows the definition of a perturbation, ${\cal Z}$, that is conserved on superhorizon scales in the absence of non-adiabatic perturbations \cite{Wands_etal_2000, Cardoso_Wands, Bertschinger2006}. In the literature ${\cal Z}$ is more commonly denoted as $\zeta$, but the choice of notation for one of the MGFs in this paper and previous work  prevents us from reusing that letter. ${\cal Z}$ is identified with the curvature perturbation on uniform-expansion hypersurfaces in a homogeneous and isotropic spacetime, which in GR coincide with hypersurfaces of constant energy density. For zeroth-order Einstein equations of the form: 
\begin{eqnarray}
{\cal H}^2&=&\frac{a^2}{3}f_0\nonumber\\
\dot{\cal H}-{\cal H}^2&=&-\frac{a^2}{2}g_0
\label{zero_order_fg}
\end{eqnarray}
the conserved perturbation is \cite{Cardoso_Wands}:
\begin{equation}
\label{zeta_definition}
{\cal Z}=-\Phi-\frac{\cal H}{\dot f_0}\delta f
\end{equation}
It would be interesting to know how ${\cal Z}$ behaves in the metric theories considered in \textsection\ref{section:delta}. 
Since the new non-GR terms do not really originate from perturbations to a fluid, it is not immediately obvious whether they will be equivalent to adiabatic or non-adiabatic pressure perturbations. Therefore the conservation of ${\cal Z}$ does not necessarily follow.\newline

Using the linearly perturbed versions of eqns.({\ref{zero_order_fg}}), one can derive an equation for the evolution of the metric potentials \cite{Cardoso_Wands}:
\begin{equation}
\label{ode}
\ddot\Phi+\frac{3{\cal H}\dot{\cal H}-\ddot{\cal H}-{\cal H}^3}{\dot{\cal H}-{\cal H}^2}\dot\Phi+{\cal H}\dot\Psi+\frac{2\dot{\cal H}^2-{\cal H}\ddot{\cal H}}{\dot{\cal H}-{\cal H}^2}\Psi=\frac{a^2}{2}\delta g_{nad}
\end{equation}
where the perturbation $\delta g$ has been decomposed into parts equivalent to adiabatic and non-adiabatic pressure perturbations:
\begin{equation}
\label{delta_g}
\delta g=\frac{\dot g_0}{\dot f_0}\delta f+\delta g_{nad}
\end{equation}
The time derivative of ${\cal Z}$ is related to the quantity on the LHS of eqn.(\ref{ode}). Hence the rate of change of ${\cal Z}$ is found to be:
\begin{equation}
\label{zeta_dot}
\dot{\cal Z}=\frac{\cal H}{\dot{\cal H}-{\cal H}^2}\frac{a^2}{2}\delta g_{nad}
\end{equation}
Comparing eqns.(\ref{einstein}) and (\ref{zero_order_fg}) and defining \mbox{$U_{00}=X$}, \mbox{$U_{ii}=Y$} we can read off:
 \begin{eqnarray}
 f_0&=&\kappa \rho_M+X\label{f_0}\nonumber\\
 g_0&=&\kappa (\rho_M+P_M)+X+Y\label{g_0}
 \end{eqnarray}
Rearranging eq.(\ref{delta_g}) and substituting in our expressions for $f$ and $g$, one finds:
\begin{eqnarray}
\hspace{-0.3cm}\delta g_{nad}&=&\frac{\kappa\dot\rho_M\dot X}{\kappa\dot\rho_M+\dot X}\left(c_s^{2(M)}-c_s^{2(X)}\right)\Gamma_{\rho_M X}+\kappa \delta P_{nad}+\delta Y_{nad}\nonumber\\
\label{delta_gnad}
\end{eqnarray}
where 
\begin{eqnarray}
\Gamma_{\rho_M X}&=&\frac{\delta\rho_M}{\dot\rho_M}-\frac{\delta X}{\dot X}\nonumber\\
c_s^{2(M)}&=&\frac{\dot P_M}{\dot\rho_M}\nonumber\\
c_s^{2(X)}&=&\frac{\dot Y}{\dot X}\nonumber
\end{eqnarray}
The perturbations $\delta P_M$ and $\delta Y$ have been decomposed in a manner analogous to eq.(\ref{delta_g}). $\Gamma_{\rho_M X}$ represents a possible entropy perturbation between CDM and the modified sector, which can be non-zero even if each component  does not support entropy perturbations by itself. 
Perturbations of the background quantities $X$ and $Y$ correspond to components of the tensor $U_{\mu\nu}$: \mbox{$a^2 \delta X=-a^2 \delta U_0^0$} and \mbox{$a^2\delta Y=a^2\delta U^i_i/3$}. Then the effective non-adiabatic pressure perturbation of the modified sector is:
\begin{equation}
\label{Y_nad}
\delta Y_{nad}=\delta Y-c_s^{2(X)}\delta X=\frac{\delta U_i^i}{3}+c_s^{2(X)}\delta U_0^0
\end{equation}
Our toy model of an EdS universe contains only cold dark matter, so \mbox{$\delta P_M=\delta P_{nad}=c_s^{2(M)}=0$}. 

\subsection{Metric-Only Theories}
For a purely metric theory with a $\Lambda$CDM-like background $X$ and $Y$ are zero or equivalent to a cosmological constant  (i.e. \mbox{$X+Y=0$}), so there can be no entropy perturbations \textit{between} CDM and the modifications (since \mbox{$c_s^{2(X)}=0$}). The only possible source for non-conservation of $\cal Z$ would be from non-adiabatic perturbations within the modified sector itself, $\delta Y_{\mathrm{nad}}$. Eqn.(\ref{zeta_dot}) becomes:
\begin{eqnarray}
\label{zeta_dot_metric}
\dot{\cal Z}&=&\frac{\cal H}{2(\dot{\cal H}-{\cal H}^2)}\frac{a^2\delta U_i^i}{3}\nonumber\\
&=&\frac{\cal H}{6(\dot{\cal H}-{\cal H}^2)}\left(k^2C_0\Phi+kC_1\dot\Phi\right) 
\end{eqnarray}
Non-adiabatic perturbations within the modified sector would amount to fluctuations about $\omega_E=-1$, which could lead to a situation equivalent to a phantom field. Whilst a phantom equation of state is permitted  by current data \cite{WMAP7_Komatsu, Nesseris_Perivo,Perivolaropoulos}, direct phantom scalar field models are plagued by severe difficulties because they lead to an unstable vacuum state \cite{Cline_2004} and favour an anisotropic universe \cite{Gannouji_Polarski}. However, it is known that theories such as scalar-tensor gravity, $f(R)$ gravity and some Lorentz-violating models can cause phases of an effective $\omega_E<-1$ without introducing a phantom field \textit{per se} \cite{Abdalla_Nojiri_Odintsov, Amendola_Tsujikawa, Martin_Schimd_Uzan, Libanov_Rubakov, Kunz_Sapone_phantom}.

Fortunately the question of whether non-adiabatic perturbations lead to \mbox{$\omega_E<-1$} turns out to be a moot point for the theories considered in \textsection\ref{section:param_choice} and \textsection\ref{section:delta}. This is because the effective pressure perturbation in eqn.(\ref{Y_nad}) vanishes on very large scales. The functions $C_0$ and $C_1$ are related to $\mu_P$ and $\zeta$ by a set of constraint equations, displayed in appendix \ref{app:Z_consv}. There we also show that under the assumptions mades in this paper \mbox{$\lim_{k\to 0}\left(k^2 C_0\right)=0$} and \mbox{$\lim_{k\to 0}\left(k C_1\right)=0$} in both of the parameterizations described in \textsection\ref{section:param_choice} . These results agree with the conclusions of \cite{Pogosian_Silvestri}.

\subsection{Theories with Additional Degrees of Freedom}
For theories with extra scalar degrees of freedom the situation is different. Firstly, the final term in $\delta Y_{nad}$ does not vanish, as the modifications to the background equations lead to $c_s^{2(X)}\neq 0$. Secondly, we cannot derive a system of constraint equations on the functions $A_0\ldots K_1$ like those in Table \ref{tab:2nd_order_constraints}, because the new scalar now acts as a source in the Bianchi identities (see appendix \ref{app:Z_consv} for details). Thirdly, we can now have entropy perturbations between matter and the modified sector, as the coefficient of $\Gamma_{\rho_M X}$ in eqn.(\ref{delta_gnad}) no longer vanishes. 

Maintaining the approach of \textsection\ref{section:extra_dof}, we keep the metric components of $\delta U_{\mu\nu}$ distinct, but treat the perturbations of the new d.o.f. as an effective fluid. We previously set the equation of state of this effective fluid to zero in order to preserve the EdS nature of our toy model; let us restore the general case for the present. The full expression for $\dot{\cal Z}$ then becomes:
\begin{widetext}
\begin{eqnarray}
\label{zeta_dot_general}
\dot{\cal Z}&=&\frac{\cal H}{2\left(\dot{\cal H}-{\cal H}^2\right)}\Bigg[\frac{\kappa \dot{\rho_M}\dot{X}}{\kappa\dot{\rho}_M+\dot{X}}\,c_s^{2(X)}\left(\frac{a^2\delta_M}{3\adotoa}+\frac{\kappa a^2 \rho_E\delta_E}{\dot X}\right)+\kappa a^2 \rho_E\left(\Pi_E-c_s^{2(X)}\delta_E\right)\nonumber\\
&&+k^2\Phi\left(\frac{1}{3}C_0-A_0\,c_s^{2(X)}\frac{\dot X}{\kappa\dot{\rho}_M+\dot{X}}\right)+k^2\Gamma\left(\frac{1}{3}J_0-F_0\, c_s^{2(X)}\frac{\dot X}{\kappa\dot{\rho}_M+\dot{X}}\right)+\frac{1}{3}k\left(C_1\dot\Phi+J_1\dot\Gamma\right)\Bigg]
\end{eqnarray}
\end{widetext}

If the MGFs do not contain inverse powers of $k$ then the second line vanishes on very large scales, leaving only perturbations to the effective fluid  (although a similar assumption was used in \textsection\ref{section:extra_dof} to make an analytic solution achievable, it does not necessarily have to hold true for all theories).  
The first set of round brackets represents entropy perturbations between CDM and the modified sector, and the second term is akin to non-adiabatic perturbations within the modified sector itself. However, note that $c_s^{2(X)}$ is not necessarily equal to the sound speed of the effective fluid, since $X$ and $Y$ may contain terms constructed from the metric. This feature distinguishes a dark energy model like quintessence from a modified gravity model such as a scalar-tensor gravity. In the latter the scalar field is non-trivially coupled to the metric and hence $X$ and $Y$ contain more terms than just the energy density and pressure of a scalar field. A quintessence-like case can be recovered from the above expression by setting the MGFs to zero and \mbox{$X=\kappa\rho_E$}, \mbox{$Y=\kappa\,P_E$}, which gives:
\begin{eqnarray}
\hspace{-1cm}
\dot{\cal Z}&=&-\frac{\cal H}{\left(\dot{\cal H}-{\cal H}^2\right)}\frac{a^2}{2}\Bigg[\frac{\rho_E(1+\omega_E)\,c_s^{2(E)}\Gamma_{M,E}}{1+(1+\omega_E)\frac{\Omega_E}{\Omega_M}}+\kappa\, \delta P_{E, \mathrm{nad}}\Bigg]\nonumber\\
\end{eqnarray}
where
\begin{equation}
\Gamma_{M,E}=\delta_M-\frac{\delta_E}{1+\omega_E}
\end{equation}
If the quintessence field supports only adiabatic perturbations ($\delta P_{E,\mathrm{nad}}=0$) then $\Gamma_{M,E}$ can be set to zero through choice of adiabatic initial conditions, leaving ${\cal Z}$ conserved.

As a second example, consider the theory treated in \textsection\ref{section:extra_dof} in which the effective fluid was assumed to be pressureless. Eqn.(\ref{zeta_dot_general}) reduces to:
\begin{eqnarray}
\dot{\cal Z}&=&-\frac{\cal H}{\left(\dot{\cal H}-{\cal H}^2\right)}\frac{\kappa a^2}{2}\frac{\dot{X}}{\left(\kappa\dot{\rho}_M+\dot{X}\right)}\,c_s^{2(X)}\left[\rho_M\delta_M+\rho_E\delta_E\right]\nonumber\\
\end{eqnarray}
The square brackets can be set to zero at through a choice of isocurvature initial conditions, so $\dot{\cal Z}$ remains conserved if $\omega_E$ is constant. However, as one wishes the effects of modified gravity to become apparent at late times in the universe an evolving $\omega_E$ would be more desirable, for which $\dot{\cal Z}$ would not be conserved.

\section{Conclusions}
\label{section:discussion}
Awareness of a forthcoming `data deluge' from current and future cosmological experiments has led to an interest in model-independent approaches to constraining modified gravity. By exploiting the generic features of current models, these parameterized systems seek to constrain large regions of theory space simultaneously. Whilst there has been much effort made to constrain these parameterizations with the data \cite{Bean, DanielLinder, Zhao2010}, there has been relatively little investigation into the corresponding theoretical description: how does the parameterized system of perturbation equations evolve? (Although see \cite{Ferreira_Skordis}; general scalar field-type models are treated in \cite{Tsujikawa_2007,deFelice_2010,Brax_Davis}).

We have attempted to answer this question within the simplified setting of an Einstein-de Sitter universe, for two classes of theories: 1) those for which the degrees of freedom are the metric and matter perturbations, and 2) theories which explicitly introduce additional scalar degrees of freedom. In both cases we have found that the curvature perturbation remains constant on the very largest scales (well above our observable horizon).

On subhorizon scales we found that in the metric-only case perturbations grow in familiar power-law fashion, but the exponents are modified from their General Relativistic values. The two modified gravity functions $\mu_P$ (controlling the effective Newton's constant) and $\zeta$ (controlling the dominant component of the gravitational slip) are partially degenerate in their effects, making them difficult to disentangle. The impact of $\zeta$ becomes subdominant if the effective Newton's constant is weakened. Modifications to the evolution of $\Phi$ lead to an induced ISW effect and modification to the growth rate, $f(z)$. 

Theories with additional scalar degrees of freedom are considerably harder to study, as the number of undetermined functions is much larger in this case. In \textsection\ref{section:extra_dof} we considered phases during which the new terms in the Einstein equations behave as an effective fluid with a negligible equation of state. We found that the modified terms dominate the evolution equations, leading to damped oscillatory or exponential behaviour. From a study of these equations we have found three relations (eqns.(\ref{constraints})) between the modified terms that, if satisfied, will restore power-law behaviour. This analysis does not apply to all modified gravity theories (when $\omega_E$ is not small), but it is relevant to (for example) EdS regimes of linear Einstein-Aether theory, EBI gravity and Ho\v{r}ava-Lifshitz gravity.

Constraint equations such as this are desirable because the number of free functions required to parameterize common gravitational theories increases rapidly when new degrees of freedom are included. This proliferation could reduce our ability to constrain such parameterized frameworks satisfactorily. One can reduce this freedom somewhat by treating the new scalar degrees of freedom as an effective fluid; indeed this (or an equivalent approach) is necessary if the parameterization is to capture theories which introduce more than two new scalar degrees of freedom. The metric perturbations are also a considerable source of freedom, as displayed in eqns.(\ref{2nd_order_Us}). This splitting of the modifications into metric parts and effective fluid parts is the key to distinguishing between closely related models of dark energy and modified gravity, such as quintessence and scalar-tensor theories.

Under the assumptions made in this paper the metric terms become irrelevant on ultra-large scales. In a metric-only theory this leaves the superhorizon perturbations ${\cal Z}$ (more commonly denoted by by $\zeta$) conserved, but in class 2) theories the possible non-conservation of $\cal Z$ depends on the equation of state of the effective fluid.

The EdS model considered in this paper is useful in obtaining a qualitative understanding of how parameterized gravity might effect the matter-dominated phase of our universe. To obtain more quantitive predictions this must be embedded in a more complex cosmological model, which is likely to be achievable only numerically.\newline

\section*{Acknowledgements}
\noindent We acknowledge useful discussions with E. Bertschinger, P. Ferreira, J. Pearson, C. Skordis, D. Wands and J. Zuntz. This work was supported by the STFC.

\appendix

\section{System of Equations for Theories with Extra D.o.F.}
\label{app:dof_eqns}
This appendix displays the system of six equations that is solved in \textsection\ref{section:extra_dof}. These are: the two spatial components of the Einstein equation (longitudinal and transverse tracless), the fluid conservation equations for cold dark matter (the two components of \mbox{$\delta\left(\nabla_{\mu}T^{\mu}_{\nu}\right)=0$}) and the two Bianchi identities \mbox{$\delta\left(\nabla_{\mu}U^{\mu}_{\nu}\right)=0$}. The non-standard terms containing MGFs arise from the parameterization laid out in eqns.(\ref{einstein}) and (\ref{2nd_order_Us}), which treats the additional scalar degrees of freedom as a pressureless fluid (denoted by a subscript $E$). This reduces to parameterization A used in \textsection\ref{section:delta} for purely metric theories.

The variable $\Psi$ in these equations can be eliminated in favour of $\Gamma$ using the definition \mbox{$\Gamma=(\dot\Phi+\adotoa\Psi)/k$}.
\begin{widetext}
\begin{eqnarray}
6k\dot\Gamma&+&12\adotoa k\Gamma+2k^2\left(\Phi-\Psi\right)+6(\dot\adotoa-\adotoa^2)\Psi=C_0 k^2\Phi+C_1 k \dot\Phi+J_0 k^2\Gamma+J_1 k \dot\Gamma\label{Einstein3}\\
\Phi-\Psi&=&D_0\Phi+\frac{D_1}{k}\dot\Phi+K_0\Gamma+\frac{K_1}{k}\dot\Gamma\label{Einstein4}\\
\dot{\delta}_M&=&-k^2\theta_M+3\dot\Phi\label{matter1}\\
\dot{\theta}_M&=&-\adotoa\theta_M+\Psi\label{matter2}\\
\kappa a^2 \rho_E \dot{\delta}_E&=&\kappa a^2 \rho_E \left(-k^2\theta_E+3\dot\Phi\right)-k^2\Phi\left(\adotoa A_0+k B_0 +\adotoa C_0\right)+k\dot\Phi\left(kA_0+\adotoa C_1\right)+k^2\Gamma\left(\adotoa F_0+k I_0 +\adotoa J_0\right)\nonumber\label{eff_fluid1}\\
&&+k\dot\Gamma\left(k F_0+\adotoa J_1\right)\\
\kappa a^2 \rho_E \dot{\theta}_E&=&\kappa a^2 \rho_E \left(-\adotoa\theta_E+\Psi\right)-k^2\Phi\left(2\frac{\adotoa}{k} B_0-\frac{1}{3} C_0 +\frac{2}{3}D_0\right)-k\dot\Phi\left(B_0-\frac{1}{3}C_1+\frac{2}{3}D_1\right)\nonumber\\
&&-k^2\Gamma\left(2\frac{\adotoa}{k} I_0-\frac{1}{3} J_0+\frac{2}{3}K_0\right)-k\dot\Gamma\left(I_0-\frac{1}{3} J_1+\frac{2}{3}K_1\right)\label{eff_fluid2}
\end{eqnarray}
\end{widetext}

\section{Conservation of ${\cal Z}$ in Metric-Only Theories}
\label{app:Z_consv}

In this appendix we demonstrate that any effective non-adiabatic perturbations that might prevent ${\cal Z}$ from being conserved vanish on large scales for the metric-only theories considered in \textsection\ref{section:param_choice} and \textsection\ref{section:delta} . First we must introduce some more detail about the parameterization underlying eqns.(\ref{gen_Poisson}) and (\ref{gen_slip}).

Consider a modified Einstein equation of the form (\ref{einstein}). In \cite{Baker2011} we demonstrated that in parameterization A (in which $\mu_s=\mu_P$) a theory containing up to second-order time derivatives corresponds to the following forms for the components of $\delta U_{\mu\nu}$ (see eqns.(\ref{U_cmpt_def}) for definitions of quantities on the left-hand side):
 \begin{eqnarray}
 \label{2nd_order_Us_metric_only}
U_{\Delta} &=& A_0 k^2\hat\Phi\nonumber\\ 
U_{\Theta} &=& B_0 k\hat\Phi\nonumber \\
U_P &=& C_0 k^2\hat\Phi + C_1 k\dot{\hat\Phi}  \nonumber\\
U_{\Sigma} &=&  D_0 \hat\Phi+\frac{D_1}{k} \dot{\hat\Phi}
\end{eqnarray}
where the coefficients $A_0,\dots D_1$ are functions of background quantities, dependencies which we will suppress to avoid cluttered expressions. $\hat\Phi$ is a gauge-invariant perturbation variable, equivalent to the Bardeen variable $-\Psi_H$, which reduces to $\Phi$ in the conformal Newtonian gauge. 
\begin{table}
\begin{tabular}{| c | l |}
\hline
& \ \qquad\; Constraint equation  \\ \hline
1 &  $\dot A_0+{\cal H} A_0+k B_0+{\cal H} C_0 = 0$\\ \hline
2 & $A_0+{\cal H}_k C_1 = 0 $\\ \hline
3 & $\dot B_0+2{\cal H} B_0-\frac{1}{3}k C_0+\frac{2}{3}k D_0 = 0$\\ \hline
4 & $B_0-\frac{1}{3} C_1+\frac{2}{3} D_1= 0$\\ \hline
\end{tabular}
\caption{Constraint equations for a metric theory in parameterization A, specified by eqns.(\ref{2nd_order_Us_metric_only}).}
\label{tab:2nd_order_constraints}
\end{table}

The two components of  \mbox{$\delta\left(\nabla_{\mu}U^{\mu}_{\nu}\right)=0$} result in equations containing $\Phi$ and its derivatives. To avoid contradicting the solution for $\Phi$ dictated by the Einstein equations these expressions must vanish identically. Setting the coefficients of each metric perturbation to zero leads to the four constraint equations listed in Table \ref{tab:2nd_order_constraints}. This enables us to reduce the six free functions in eqns.(\ref{2nd_order_Us_metric_only}) to just two. Note that terms in $\hat\Psi$ (equivalent to the Bardeen variable $\Phi_A$ ) are not permitted to appear in $\delta U^0_0$ and $\delta U^0_i$, because they contain second-order time derivatives, which would lead to third-order Bianchi identities. Constraint equations similar to those in Table \ref{tab:2nd_order_constraints} prevents $\hat\Psi$ from featuring in $\delta U^i_i$ and $\delta U^i_j$ if it is not present in $\delta U^0_0$ and $\delta U^0_i$.

We will choose one  of our free functions to be $D_0$ --  this corresponds to $\zeta$ in eqn.(\ref{gen_slip}). The second free function will be the combination that appears when we form the Poisson equation:
\begin{equation}
\mu_P=\frac{1}{1+\frac{1}{2}\left(A_0+3{\cal H}_k B_0\right)}
\end{equation}
where $\adotoa_k=\adotoa/k$.
In terms of these two MGFs the other coefficient functions are:
\begin{eqnarray}
\label{SF_coeffs}
A_0&=&-2\left(\frac{\mu_P-1}{\mu_P}\right)\left(1+\frac{{\cal H}^2}{Q}\right) +2\zeta\frac{{\cal H}^2}{Q}\label{A_0}\nonumber\\
B_0&=&\frac{2{\cal H}k\left(\frac{(\mu_P-1)}{\mu_P}-\zeta\right)}{3Q}\label{B_0}\nonumber\\
C_0&=&\frac{2}{Q}\left(\frac{(\mu_P-1)}{\mu_P}-\zeta\right)\left(\dot{\cal H}-{\cal H}\frac{\dot Q}{Q}+2{\cal H}^2\right)+2\zeta\nonumber\\
&=&\frac{4}{3}\left(\frac{(\mu_P-1)}{\mu_P}-\zeta\right)\frac{\left(\frac{5}{2}+\frac{1}{3{\cal H}_k^2}\right)}{\left(1+\frac{2}{9{\cal H}_k^2}\right)^2}+2\zeta\nonumber\\
C_1&=&\frac{2}{{\cal H}_k}\left(\frac{\mu_P-1}{\mu_P}\right)\left(1+\frac{{\cal H}^2}{Q}\right)-2\frac{{\cal H}k}{Q}\zeta\label{C_1}\nonumber\\
D_1&=&\frac{1}{{\cal H}_k}\left(\frac{\mu_P-1}{\mu_P}\right)\nonumber\\
\mathrm{where}&&Q={\cal H}^2+\frac{k^2}{3}-\dot{\cal H}.
\end{eqnarray}
The second expression for $C_0$ applies only in an EdS universe. From eqn.(\ref{zeta_dot_metric}) we see that the quantity of interest for evaluating $\dot{\cal{Z}}$ is $\delta U^i_i$. In fact we only need to know $C_0$, since we showed in \textsection\ref{sub:superhorizon_scales} that the potential is constant on large scales.
Under the assumption that $\mu_P$ and $\zeta$ do not contain inverse powers of $k$, eqns.(\ref{2nd_order_Us_metric_only}) and (\ref{SF_coeffs}) imply that \mbox{$\lim_{k\to 0} \delta U_i^i=0$}.  Hence ${\cal Z}$ is conserved on superhorizon scales in parameterization A.\newline

\begin{table}[t!]
\begin{tabular}{| c | l |}
\hline
& \qquad\qquad Constraint equation  \\ \hline
1 & $\dot A_0+{\cal H}A_0+k B_0+{\cal H}C_0=0$ \\ \hline
2 & $\dot A_1+{\cal H}A_1+k A_0+k B_1+{\cal H}C_1=0$ \\ \hline
3 & $k A_1+{\cal H}C_2=0$ \\ \hline
4 & $\dot B_0+2{\cal H}B_0-\frac{1}{3}kC_0+\frac{2}{3}k D_0=0$ \\ \hline
5 & $\dot B_1+2{\cal H}B_1+k B_0-\frac{1}{3}k C_1=0$ \\ \hline
6 & $ B_1-\frac{1}{3}C_2=0$ \\ \hline
\end{tabular}
\caption{Table of the constraint equations for a metric theory in parameterization B, specified by eqns.(\ref{U_3rd_order}).}
\label{tab:3rd_order_constraints}
\end{table}

We can repeat this calculation for parameterization B, in which $\mu_s=1$ but $\mu_P$ remains a free function. This corresponds to a $\delta U_{\mu\nu}$ tensor of the following form (see \cite{Baker2011}):
 \begin{eqnarray}
 \label{3rd_order_Us}
U_{\Delta} &=& A_0 k^2\hat\Phi+A_1 k\dot{\hat\Phi}\nonumber\\
U_{\Theta}&=& B_0 k\hat\Phi+B_1\dot{\hat\Phi}\nonumber \\
U_P &=& C_0 k^2\hat\Phi + C_1 k\dot{\hat\Phi}+C_2 \ddot{\hat\Phi}  \nonumber\\
U_{\Sigma} &=&  D_0 \hat\Phi \label{U_3rd_order}
\end{eqnarray}
The constraint equations for this theory are listed in Table \ref{tab:3rd_order_constraints}. In terms of the MGFs $\mu_P$ and $\zeta$ the coefficients are:
\begin{eqnarray}
A_0&=&-2\left(\frac{\mu_P-1}{\mu_P}\right)\left(1+\frac{{\cal H}^2}{Q}\right) +2\frac{{\cal H}^2}{Q}\zeta\label{A_0}\nonumber\\
A_1&=&\frac{-6{\cal H}_k}{1+3{\cal H}_k^2}\left(\frac{\mu_P-1}{\mu_P}\right)\nonumber\\
B_0&=&\frac{2{\cal H}k\left(\frac{(\mu_P-1)}{\mu_P}-\zeta\right)}{3Q}\label{B_0}\nonumber\\
B_1&=&\frac{2}{(1+3{\cal H}_k^2)}\left(\frac{\mu_P-1}{\mu_P}\right)\nonumber\\
C_0&=&\frac{2}{Q}\left(\frac{(\mu_P-1)}{\mu_P}-\zeta\right)\left(\dot{\cal H}-{\cal H}\frac{\dot Q}{Q}+2\adotoa^2\right)+2\zeta\nonumber\\
&=&\frac{2}{Q}\left(\frac{(\mu_P-1)}{\mu_P}-\zeta\right)\left(\frac{3}{2}{\cal H}^2-{\cal H}\frac{\dot Q}{Q}\right)+2\zeta\nonumber
\end{eqnarray}
\begin{eqnarray}
C_1&=&\frac{2\adotoa k}{Q}\left(\frac{(\mu_P-1)}{\mu_P}-\zeta\right)\nonumber\\
&&+4\adotoa k\left(\frac{(\mu_P-1)}{\mu_P}\right)\frac{\left(Q+\frac{2}{3}\dot\adotoa\right)}{\left(Q+\dot\adotoa\right)^2}\nonumber\\
&=&\frac{2\adotoa k}{Q}\left(\frac{(\mu_P-1)}{\mu_P}-\zeta\right)\nonumber\\
&&+4\adotoa k\left(\frac{(\mu_P-1)}{\mu_P}\right)\frac{\left(Q-\frac{1}{3}\adotoa^2\right)}{\left(Q-\frac{1}{2}\adotoa^2\right)^2}\nonumber\\
C_2&=&\frac{6}{(1+3{\cal H}_k^2)}\left(\frac{\mu_P-1}{\mu_P}\right)
\end{eqnarray}
As we found for the parameterization A case, $\lim_{k\to 0}\delta U^i_i=0$. Hence ${\cal Z}$ is conserved in parameterization B also.

\section{$\delta_M$ and $\delta_E$ in Theories with Extra D.o.F.}
\label{app:deltas_in_dof}

We wish to connect the subhorizon solution for $\Phi$ in eqn.(\ref{phi_sol_dof_sub}) to the density perturbations of CDM and the effective fluid.
First we cast eqns.(\ref{Einstein3}) and (\ref{Einstein4}) in dimensionless format by using the substitution $x=k\eta$ (and correspondingly $\adotoa=2/x$, $\dot\adotoa=-2/x^2$). We also apply the conditions necessary for power-law growth, given in eqns.(\ref{constraints}). The spatial components of the Einstein equations become (where primes denote derivatives with respect to $x$):
\begin{eqnarray}
&&\Gamma^{\prime}(6-J_1)+\Gamma \left(\frac{6}{x}-x-2K_0\right)+\Phi^{\prime}\left(\frac{18}{x}+x-2D_1\right)\nonumber\\
&&\hspace{5.4cm}+\,2\Phi (1-D_0)=0\nonumber\\
&&\Gamma^{\prime} K_1+\Gamma \left(K_0+\frac{x}{2}\right)+\Phi^{\prime}\left(D_1-\frac{x}{2}\right)-\Phi(1-D_0)=0\nonumber
\end{eqnarray}
Eliminating $\Gamma^{\prime}$ from these:
\begin{eqnarray}
\label{appC_phi_prime}
\Phi^{\prime}\left[\frac{18}{x}+\frac{\gamma}{K_1}\left(\frac{x}{2}-1\right)\right]&+&\Phi\frac{\gamma}{K_1}(1-D_0)\\
&+&\Gamma\left[\frac{6}{x}-\frac{\gamma}{K_1}\left(\frac{x}{2}+1\right)\right]=0\nonumber
\end{eqnarray}
where $\gamma=6-J_1+2K_1$. Keeping only the growing mode, we write the solution for $\Phi$ as:
\begin{equation}
\Phi(x)=R(k) x^{q}
 \end{equation}
We have absorbed the $k$-dependence into the prefactor, $R(k)=P/k^q$ where $P$ is a constant (see eqn.(\ref{phi_sol_dof_sub})), and dropped the superscript $+$ on $q$ to avoid cluttered expressions.

Taking the subhorizon limit $x\to\infty$, eqn.(\ref{appC_phi_prime}) gives us the following solution for $\Gamma$:
\begin{equation}
\Gamma(x)=2 R(k)\left(\frac{q}{2}+1-D_0\right)x^{q-1}
\end{equation}
Next we recast the fluid conservation equations for CDM in dimensionless format and combine them in a manner analogous to eqns.(\ref{fluid_eqn1})-(\ref{delta_evol_2}), yielding:
\begin{equation}
\delta_M^{\prime\prime}+\frac{2}{x}\delta_M^{\prime}=3\Phi^{\prime\prime}+\Phi^{\prime}\left(\frac{6}{x}+\frac{x}{2}\right)-\frac{x}{2}\Gamma
\end{equation}
This has the solution:
\begin{equation}
\delta_M (x)=c_1+\frac{c_2}{x}+R(k)x^q\left[\frac{(D_0-1)x^2}{(q+2)(q+3)}+3\right]
\end{equation}
where $c_1$ and $c_2$ are integration constants. This solution has the desired behaviour that when we set $D_0=0$ (which sets $q=0$, see eqn.(\ref{phi_sol_dof_sub})) we recover that $\delta_M$ grows as $\delta_M\propto x^2\propto a$, as occurs in GR-controlled matter-dominated epoch. Finally we use the (dimensionless) Poisson equation to relate the solutions for $\Phi (x)$ and $\delta_M (x)$ to $\delta_E (x)$:
\begin{equation}
\label{dimless_dof_Poisson}
-2\Phi=\frac{12\mu_P}{x^2}\left(\Omega_M\delta_M+\Omega_E\delta_E\right)+\Gamma\mu_P\left(F_0+\frac{6}{x} I_0\right)
\end{equation} 
In our toy EdS universe, containing only CDM and the effective fluid, $\Omega_M$ and $\Omega_E$ are constants. Retaining just the dominant growing modes for both $\delta_M (x)$ and $\delta_E (x)$, the above equation gives us the result:
\begin{equation}
\label{delta_E_sol}
\delta_E (x)=\frac{R(k)}{6\Omega_E}\left[\frac{6(1-D_0)(1-\Omega_E)}{(q+2)(q+3)}-\frac{1}{\mu_P}\right]x^{q+2}
\end{equation}
As expected, $\delta_M (x)$ and $\delta_E (x)$ grow at the same rate. Note that, as in GR, their evolution differs from that of $\Phi (x)$ by a single factor of $a$.


\bibliographystyle{apsrev}

\begin{thebibliography}{78}
\expandafter\ifx\csname natexlab\endcsname\relax\def\natexlab#1{#1}\fi
\expandafter\ifx\csname bibnamefont\endcsname\relax
  \def\bibnamefont#1{#1}\fi
\expandafter\ifx\csname bibfnamefont\endcsname\relax
  \def\bibfnamefont#1{#1}\fi
\expandafter\ifx\csname citenamefont\endcsname\relax
  \def\citenamefont#1{#1}\fi
\expandafter\ifx\csname url\endcsname\relax
  \def\url#1{\texttt{#1}}\fi
\expandafter\ifx\csname urlprefix\endcsname\relax\def\urlprefix{URL }\fi
\providecommand{\bibinfo}[2]{#2}
\providecommand{\eprint}[2][]{\url{#2}}

\bibitem[{\citenamefont{{Nicolis} et~al.}(2009)\citenamefont{{Nicolis},
  {Rattazzi}, and {Trincherini}}}]{Nicolis_Galileons}
\bibinfo{author}{\bibfnamefont{A.}~\bibnamefont{{Nicolis}}},
  \bibinfo{author}{\bibfnamefont{R.}~\bibnamefont{{Rattazzi}}},
  \bibnamefont{and}
  \bibinfo{author}{\bibfnamefont{E.}~\bibnamefont{{Trincherini}}},
  \bibinfo{journal}{\prd} \textbf{\bibinfo{volume}{79}},
  \bibinfo{pages}{064036} (\bibinfo{year}{2009}), \eprint{0811.2197}.

\bibitem[{\citenamefont{{Rubakov}}(2001)}]{Rubakov}
\bibinfo{author}{\bibfnamefont{V.~A.} \bibnamefont{{Rubakov}}},
  \bibinfo{journal}{Physics Uspekhi} \textbf{\bibinfo{volume}{44}},
  \bibinfo{pages}{871} (\bibinfo{year}{2001}), \eprint{arXiv:hep-ph/0104152}.

\bibitem[{\citenamefont{{Maartens} and {Koyama}}(2010)}]{MaartensKoyama}
\bibinfo{author}{\bibfnamefont{R.}~\bibnamefont{{Maartens}}} \bibnamefont{and}
  \bibinfo{author}{\bibfnamefont{K.}~\bibnamefont{{Koyama}}},
  \bibinfo{journal}{Living Reviews in Relativity}
  \textbf{\bibinfo{volume}{13}}, \bibinfo{pages}{5} (\bibinfo{year}{2010}),
  \eprint{1004.3962}.

\bibitem[{\citenamefont{{Clifton} et~al.}(2011)\citenamefont{{Clifton},
  {Ferreira}, {Padilla}, and {Skordis}}}]{MG_report}
\bibinfo{author}{\bibfnamefont{T.}~\bibnamefont{{Clifton}}},
  \bibinfo{author}{\bibfnamefont{P.~G.} \bibnamefont{{Ferreira}}},
  \bibinfo{author}{\bibfnamefont{A.}~\bibnamefont{{Padilla}}},
  \bibnamefont{and}
  \bibinfo{author}{\bibfnamefont{C.}~\bibnamefont{{Skordis}}},
  \bibinfo{journal}{ArXiv e-prints}  (\bibinfo{year}{2011}),
  \eprint{1106.2476}.

\bibitem[{\citenamefont{{Nojiri} and
  {Odintsov}}(2011)}]{Nojiri_Odintsov_review}
\bibinfo{author}{\bibfnamefont{S.}~\bibnamefont{{Nojiri}}} \bibnamefont{and}
  \bibinfo{author}{\bibfnamefont{S.~D.} \bibnamefont{{Odintsov}}},
  \bibinfo{journal}{Physics Reports} \textbf{\bibinfo{volume}{505}},
  \bibinfo{pages}{59} (\bibinfo{year}{2011}), \eprint{1011.0544}.

\bibitem[{\citenamefont{{Will}}(1971)}]{Will1971}
\bibinfo{author}{\bibfnamefont{C.~M.} \bibnamefont{{Will}}},
  \bibinfo{journal}{\apj} \textbf{\bibinfo{volume}{163}}, \bibinfo{pages}{611}
  (\bibinfo{year}{1971}).

\bibitem[{\citenamefont{{Thorne} and {Will}}(1971)}]{Thorne_Will}
\bibinfo{author}{\bibfnamefont{K.~S.} \bibnamefont{{Thorne}}} \bibnamefont{and}
  \bibinfo{author}{\bibfnamefont{C.~M.} \bibnamefont{{Will}}},
  \bibinfo{journal}{\apj} \textbf{\bibinfo{volume}{163}}, \bibinfo{pages}{595}
  (\bibinfo{year}{1971}).

\bibitem[{\citenamefont{{Will} and {Nordtvedt}}(1972)}]{Will_Nordvedt_1972}
\bibinfo{author}{\bibfnamefont{C.~M.} \bibnamefont{{Will}}} \bibnamefont{and}
  \bibinfo{author}{\bibfnamefont{K.}~\bibnamefont{{Nordtvedt}},
  \bibfnamefont{Jr.}}, \bibinfo{journal}{\apj} \textbf{\bibinfo{volume}{177}},
  \bibinfo{pages}{757} (\bibinfo{year}{1972}).

\bibitem[{\citenamefont{{Will}}(2011)}]{Will2011}
\bibinfo{author}{\bibfnamefont{C.~M.} \bibnamefont{{Will}}},
  \bibinfo{journal}{ArXiv e-prints}  (\bibinfo{year}{2011}),
  \eprint{1102.5192}.

\bibitem[{\citenamefont{{Landau}}(1946)}]{Landau}
\bibinfo{author}{\bibfnamefont{L.}~\bibnamefont{{Landau}}},
  \bibinfo{journal}{Soviet J. Phys (JETP)} \textbf{\bibinfo{volume}{116}}
  (\bibinfo{year}{1946}).

\bibitem[{\citenamefont{{Peebles}}(1982)}]{Peebles}
\bibinfo{author}{\bibfnamefont{P.~J.~E.} \bibnamefont{{Peebles}}},
  \bibinfo{journal}{ApJL} \textbf{\bibinfo{volume}{263}}, \bibinfo{pages}{L1}
  (\bibinfo{year}{1982}).

\bibitem[{\citenamefont{{Peebles}}(1980)}]{Peebles_book}
\bibinfo{author}{\bibfnamefont{P.~J.~E.} \bibnamefont{{Peebles}}},
  \emph{\bibinfo{title}{{The Large-Scale Structure of the Universe}}}
  (\bibinfo{publisher}{{Princeton University Press}}, \bibinfo{year}{1980}).

\bibitem[{\citenamefont{{Blumenthal} et~al.}(1984)\citenamefont{{Blumenthal},
  {Faber}, {Primack}, and {Rees}}}]{Blumenthal}
\bibinfo{author}{\bibfnamefont{G.~R.} \bibnamefont{{Blumenthal}}},
  \bibinfo{author}{\bibfnamefont{S.~M.} \bibnamefont{{Faber}}},
  \bibinfo{author}{\bibfnamefont{J.~R.} \bibnamefont{{Primack}}},
  \bibnamefont{and} \bibinfo{author}{\bibfnamefont{M.~J.}
  \bibnamefont{{Rees}}}, \bibinfo{journal}{\nat}
  \textbf{\bibinfo{volume}{311}}, \bibinfo{pages}{517} (\bibinfo{year}{1984}).

\bibitem[{\citenamefont{{Meszaros}}(1974)}]{Meszaros}
\bibinfo{author}{\bibfnamefont{P.}~\bibnamefont{{Meszaros}}},
  \bibinfo{journal}{A\&A} \textbf{\bibinfo{volume}{37}}, \bibinfo{pages}{225}
  (\bibinfo{year}{1974}).

\bibitem[{\citenamefont{{Nesseris} et~al.}(2011)\citenamefont{{Nesseris},
  {Blake}, {Davis}, and {Parkinson}}}]{WiggleZ_Nesseris}
\bibinfo{author}{\bibfnamefont{S.}~\bibnamefont{{Nesseris}}},
  \bibinfo{author}{\bibfnamefont{C.}~\bibnamefont{{Blake}}},
  \bibinfo{author}{\bibfnamefont{T.}~\bibnamefont{{Davis}}}, \bibnamefont{and}
  \bibinfo{author}{\bibfnamefont{D.}~\bibnamefont{{Parkinson}}},
  \bibinfo{journal}{JCAP} \textbf{\bibinfo{volume}{7}}, \bibinfo{pages}{37}
  (\bibinfo{year}{2011}), \eprint{1107.3659}.

\bibitem[{\citenamefont{{Bambi} et~al.}(2005)\citenamefont{{Bambi},
  {Giannotti}, and {Villante}}}]{Bambi_Giannotti}
\bibinfo{author}{\bibfnamefont{C.}~\bibnamefont{{Bambi}}},
  \bibinfo{author}{\bibfnamefont{M.}~\bibnamefont{{Giannotti}}},
  \bibnamefont{and} \bibinfo{author}{\bibfnamefont{F.~L.}
  \bibnamefont{{Villante}}}, \bibinfo{journal}{\prd}
  \textbf{\bibinfo{volume}{71}}, \bibinfo{pages}{123524}
  (\bibinfo{year}{2005}), \eprint{arXiv:astro-ph/0503502}.

\bibitem[{\citenamefont{{Calabrese} et~al.}(2011)\citenamefont{{Calabrese},
  {Huterer}, {Linder}, {Melchiorri}, and {Pagano}}}]{Calabrese}
\bibinfo{author}{\bibfnamefont{E.}~\bibnamefont{{Calabrese}}},
  \bibinfo{author}{\bibfnamefont{D.}~\bibnamefont{{Huterer}}},
  \bibinfo{author}{\bibfnamefont{E.~V.} \bibnamefont{{Linder}}},
  \bibinfo{author}{\bibfnamefont{A.}~\bibnamefont{{Melchiorri}}},
  \bibnamefont{and} \bibinfo{author}{\bibfnamefont{L.}~\bibnamefont{{Pagano}}},
  \bibinfo{journal}{\prd} \textbf{\bibinfo{volume}{83}},
  \bibinfo{pages}{123504} (\bibinfo{year}{2011}), \eprint{1103.4132}.

\bibitem[{\citenamefont{{Reichardt} et~al.}(2011)\citenamefont{{Reichardt}, {de
  Putter}, {Zahn}, and {Hou}}}]{SPT_EDE}
\bibinfo{author}{\bibfnamefont{C.~L.} \bibnamefont{{Reichardt}}},
  \bibinfo{author}{\bibfnamefont{R.}~\bibnamefont{{de Putter}}},
  \bibinfo{author}{\bibfnamefont{O.}~\bibnamefont{{Zahn}}}, \bibnamefont{and}
  \bibinfo{author}{\bibfnamefont{Z.}~\bibnamefont{{Hou}}},
  \bibinfo{journal}{ArXiv e-prints}  (\bibinfo{year}{2011}),
  \eprint{1110.5328}.

\bibitem[{\citenamefont{{Brax} and {Davis}}(2011)}]{Brax_Davis}
\bibinfo{author}{\bibfnamefont{P.}~\bibnamefont{{Brax}}} \bibnamefont{and}
  \bibinfo{author}{\bibfnamefont{A.-C.} \bibnamefont{{Davis}}},
  \bibinfo{journal}{ArXiv e-prints}  (\bibinfo{year}{2011}),
  \eprint{1109.5862}.

\bibitem[{\citenamefont{{Skordis}}(2009)}]{Skordis2009}
\bibinfo{author}{\bibfnamefont{C.}~\bibnamefont{{Skordis}}},
  \bibinfo{journal}{\prd} \textbf{\bibinfo{volume}{79}},
  \bibinfo{pages}{123527} (\bibinfo{year}{2009}), \eprint{0806.1238}.

\bibitem[{\citenamefont{{Ferreira} and {Skordis}}(2010)}]{Ferreira_Skordis}
\bibinfo{author}{\bibfnamefont{P.~G.} \bibnamefont{{Ferreira}}}
  \bibnamefont{and}
  \bibinfo{author}{\bibfnamefont{C.}~\bibnamefont{{Skordis}}},
  \bibinfo{journal}{\prd} \textbf{\bibinfo{volume}{81}},
  \bibinfo{pages}{104020} (\bibinfo{year}{2010}), \eprint{1003.4231}.

\bibitem[{\citenamefont{{Baker} et~al.}(2011)\citenamefont{{Baker}, {Ferreira},
  {Skordis}, and {Zuntz}}}]{Baker2011}
\bibinfo{author}{\bibfnamefont{T.}~\bibnamefont{{Baker}}},
  \bibinfo{author}{\bibfnamefont{P.~G.} \bibnamefont{{Ferreira}}},
  \bibinfo{author}{\bibfnamefont{C.}~\bibnamefont{{Skordis}}},
  \bibnamefont{and} \bibinfo{author}{\bibfnamefont{J.}~\bibnamefont{{Zuntz}}},
  \bibinfo{journal}{ArXiv e-prints}  (\bibinfo{year}{2011}),
  \eprint{1107.0491}.

\bibitem[{\citenamefont{{Zuntz} et~al.}(2011)\citenamefont{{Zuntz}, {Baker},
  {Ferreira}, and {Skordis}}}]{Zuntz2011}
\bibinfo{author}{\bibfnamefont{J.}~\bibnamefont{{Zuntz}}},
  \bibinfo{author}{\bibfnamefont{T.}~\bibnamefont{{Baker}}},
  \bibinfo{author}{\bibfnamefont{P.}~\bibnamefont{{Ferreira}}},
  \bibnamefont{and}
  \bibinfo{author}{\bibfnamefont{C.}~\bibnamefont{{Skordis}}},
  \bibinfo{journal}{ArXiv e-prints}  (\bibinfo{year}{2011}),
  \eprint{1110.3830}.

\bibitem[{\citenamefont{Hu}(1998)}]{HU_GDM}
\bibinfo{author}{\bibfnamefont{W.}~\bibnamefont{Hu}},
  \bibinfo{journal}{Astrophys. J.} \textbf{\bibinfo{volume}{506}},
  \bibinfo{pages}{485} (\bibinfo{year}{1998}), \eprint{astro-ph/9801234}.

\bibitem[{\citenamefont{Bean and Tangmatitham}(2010)}]{Bean}
\bibinfo{author}{\bibfnamefont{R.}~\bibnamefont{Bean}} \bibnamefont{and}
  \bibinfo{author}{\bibfnamefont{M.}~\bibnamefont{Tangmatitham}},
  \bibinfo{journal}{Phys. Rev.} \textbf{\bibinfo{volume}{D81}},
  \bibinfo{pages}{083534} (\bibinfo{year}{2010}), \eprint{1002.4197}.

\bibitem[{\citenamefont{Daniel and Linder}(2010)}]{DanielLinder}
\bibinfo{author}{\bibfnamefont{S.~F.} \bibnamefont{Daniel}} \bibnamefont{and}
  \bibinfo{author}{\bibfnamefont{E.~V.} \bibnamefont{Linder}},
  \bibinfo{journal}{Phys. Rev. D} \textbf{\bibinfo{volume}{82}},
  \bibinfo{pages}{103523} (\bibinfo{year}{2010}).

\bibitem[{\citenamefont{{Pogosian} et~al.}(2010)\citenamefont{{Pogosian},
  {Silvestri}, {Koyama}, and {Zhao}}}]{Pogosian_parameterization}
\bibinfo{author}{\bibfnamefont{L.}~\bibnamefont{{Pogosian}}},
  \bibinfo{author}{\bibfnamefont{A.}~\bibnamefont{{Silvestri}}},
  \bibinfo{author}{\bibfnamefont{K.}~\bibnamefont{{Koyama}}}, \bibnamefont{and}
  \bibinfo{author}{\bibfnamefont{G.-B.} \bibnamefont{{Zhao}}},
  \bibinfo{journal}{\prd} \textbf{\bibinfo{volume}{81}},
  \bibinfo{pages}{104023} (\bibinfo{year}{2010}), \eprint{1002.2382}.

\bibitem[{\citenamefont{{Hojjati}
  et~al.}(2011{\natexlab{a}})\citenamefont{{Hojjati}, {Zhao}, {Pogosian},
  {Silvestri}, {Crittenden}, and {Koyama}}}]{Hojjati_Zhao_2011}
\bibinfo{author}{\bibfnamefont{A.}~\bibnamefont{{Hojjati}}},
  \bibinfo{author}{\bibfnamefont{G.-B.} \bibnamefont{{Zhao}}},
  \bibinfo{author}{\bibfnamefont{L.}~\bibnamefont{{Pogosian}}},
  \bibinfo{author}{\bibfnamefont{A.}~\bibnamefont{{Silvestri}}},
  \bibinfo{author}{\bibfnamefont{R.}~\bibnamefont{{Crittenden}}},
  \bibnamefont{and} \bibinfo{author}{\bibfnamefont{K.}~\bibnamefont{{Koyama}}},
  \bibinfo{journal}{ArXiv e-prints}  (\bibinfo{year}{2011}{\natexlab{a}}),
  \eprint{1111.3960}.

\bibitem[{\citenamefont{{Bertschinger} and {Zukin}}(2008)}]{Bertschinger_Zukin}
\bibinfo{author}{\bibfnamefont{E.}~\bibnamefont{{Bertschinger}}}
  \bibnamefont{and} \bibinfo{author}{\bibfnamefont{P.}~\bibnamefont{{Zukin}}},
  \bibinfo{journal}{\prd} \textbf{\bibinfo{volume}{78}},
  \bibinfo{pages}{024015} (\bibinfo{year}{2008}), \eprint{0801.2431}.

\bibitem[{\citenamefont{Hu and Sawicki}(2007)}]{Hu_Sawicki}
\bibinfo{author}{\bibfnamefont{W.}~\bibnamefont{Hu}} \bibnamefont{and}
  \bibinfo{author}{\bibfnamefont{I.}~\bibnamefont{Sawicki}},
  \bibinfo{journal}{Phys. Rev.} \textbf{\bibinfo{volume}{D76}},
  \bibinfo{pages}{104043} (\bibinfo{year}{2007}), \eprint{0708.1190}.

\bibitem[{\citenamefont{{Pogosian} et~al.}(2011)\citenamefont{{Pogosian},
  {Koyama}, {Ferreira}, and {Zhao}}}]{private_comm}
\bibinfo{author}{\bibfnamefont{L.}~\bibnamefont{{Pogosian}}},
  \bibinfo{author}{\bibfnamefont{K.}~\bibnamefont{{Koyama}}},
  \bibinfo{author}{\bibfnamefont{P.}~\bibnamefont{{Ferreira}}},
  \bibnamefont{and} \bibinfo{author}{\bibfnamefont{G.}~\bibnamefont{{Zhao}}}
  (\bibinfo{year}{2011}), \bibinfo{note}{private communication}.

\bibitem[{\citenamefont{Lovelock}(1972)}]{Lovelock1}
\bibinfo{author}{\bibfnamefont{D.}~\bibnamefont{Lovelock}},
  \bibinfo{journal}{Journal of Mathematical Physics}
  \textbf{\bibinfo{volume}{13}}, \bibinfo{pages}{874} (\bibinfo{year}{1972}),
  \urlprefix\url{http://link.aip.org/link/?JMP/13/874/1}.

\bibitem[{\citenamefont{Lovelock}(1971)}]{Lovelock2}
\bibinfo{author}{\bibfnamefont{D.}~\bibnamefont{Lovelock}},
  \bibinfo{journal}{Journal of Mathematical Physics}
  \textbf{\bibinfo{volume}{12}}, \bibinfo{pages}{498} (\bibinfo{year}{1971}),
  \urlprefix\url{http://link.aip.org/link/?JMP/12/498/1}.

\bibitem[{\citenamefont{{Deser} and {Woodard}}(2007)}]{Deser_Woodard}
\bibinfo{author}{\bibfnamefont{S.}~\bibnamefont{{Deser}}} \bibnamefont{and}
  \bibinfo{author}{\bibfnamefont{R.~P.} \bibnamefont{{Woodard}}},
  \bibinfo{journal}{Physical Review Letters} \textbf{\bibinfo{volume}{99}},
  \bibinfo{pages}{111301} (\bibinfo{year}{2007}), \eprint{0706.2151}.

\bibitem[{\citenamefont{{Deffayet} and {Woodard}}(2009)}]{Deffayet_Woodard}
\bibinfo{author}{\bibfnamefont{C.}~\bibnamefont{{Deffayet}}} \bibnamefont{and}
  \bibinfo{author}{\bibfnamefont{R.~P.} \bibnamefont{{Woodard}}},
  \bibinfo{journal}{JCAP} \textbf{\bibinfo{volume}{8}}, \bibinfo{pages}{23}
  (\bibinfo{year}{2009}), \eprint{0904.0961}.

\bibitem[{\citenamefont{Bruno and Zumino}(1986)}]{Zumino}
\bibinfo{author}{\bibnamefont{Bruno}} \bibnamefont{and}
  \bibinfo{author}{\bibnamefont{Zumino}}, \bibinfo{journal}{Physics Reports}
  \textbf{\bibinfo{volume}{137}}, \bibinfo{pages}{109 } (\bibinfo{year}{1986}),
  ISSN \bibinfo{issn}{0370-1573},
  \urlprefix\url{http://www.sciencedirect.com/science/article/pii/0370157386900761}.

\bibitem[{\citenamefont{{Amendola} et~al.}(2008)\citenamefont{{Amendola},
  {Kunz}, and {Sapone}}}]{Amendola_WL}
\bibinfo{author}{\bibfnamefont{L.}~\bibnamefont{{Amendola}}},
  \bibinfo{author}{\bibfnamefont{M.}~\bibnamefont{{Kunz}}}, \bibnamefont{and}
  \bibinfo{author}{\bibfnamefont{D.}~\bibnamefont{{Sapone}}},
  \bibinfo{journal}{JCAP} \textbf{\bibinfo{volume}{4}}, \bibinfo{pages}{13}
  (\bibinfo{year}{2008}), \eprint{0704.2421}.

\bibitem[{\citenamefont{{Schimd} et~al.}(2005)\citenamefont{{Schimd}, {Uzan},
  and {Riazuelo}}}]{Schimd}
\bibinfo{author}{\bibfnamefont{C.}~\bibnamefont{{Schimd}}},
  \bibinfo{author}{\bibfnamefont{J.-P.} \bibnamefont{{Uzan}}},
  \bibnamefont{and}
  \bibinfo{author}{\bibfnamefont{A.}~\bibnamefont{{Riazuelo}}},
  \bibinfo{journal}{\prd} \textbf{\bibinfo{volume}{71}},
  \bibinfo{pages}{083512} (\bibinfo{year}{2005}),
  \eprint{arXiv:astro-ph/0412120}.

\bibitem[{\citenamefont{{Pogosian} and {Silvestri}}(2008)}]{Pogosian_Silvestri}
\bibinfo{author}{\bibfnamefont{L.}~\bibnamefont{{Pogosian}}} \bibnamefont{and}
  \bibinfo{author}{\bibfnamefont{A.}~\bibnamefont{{Silvestri}}},
  \bibinfo{journal}{\prd} \textbf{\bibinfo{volume}{77}},
  \bibinfo{pages}{023503} (\bibinfo{year}{2008}), \eprint{0709.0296}.

\bibitem[{\citenamefont{{Koyama} and
  {Maartens}}(2006)}]{KoyamaMaartens_structureDGP}
\bibinfo{author}{\bibfnamefont{K.}~\bibnamefont{{Koyama}}} \bibnamefont{and}
  \bibinfo{author}{\bibfnamefont{R.}~\bibnamefont{{Maartens}}},
  \bibinfo{journal}{JCAP} \textbf{\bibinfo{volume}{1}}, \bibinfo{pages}{16}
  (\bibinfo{year}{2006}), \eprint{arXiv:astro-ph/0511634}.

\bibitem[{\citenamefont{{Zhao} et~al.}(2009)\citenamefont{{Zhao}, {Pogosian},
  {Silvestri}, and {Zylberberg}}}]{MGCAMB}
\bibinfo{author}{\bibfnamefont{G.-B.} \bibnamefont{{Zhao}}},
  \bibinfo{author}{\bibfnamefont{L.}~\bibnamefont{{Pogosian}}},
  \bibinfo{author}{\bibfnamefont{A.}~\bibnamefont{{Silvestri}}},
  \bibnamefont{and}
  \bibinfo{author}{\bibfnamefont{J.}~\bibnamefont{{Zylberberg}}},
  \bibinfo{journal}{\prd} \textbf{\bibinfo{volume}{79}},
  \bibinfo{pages}{083513} (\bibinfo{year}{2009}), \eprint{0809.3791}.

\bibitem[{\citenamefont{{Hojjati}
  et~al.}(2011{\natexlab{b}})\citenamefont{{Hojjati}, {Pogosian}, and
  {Zhao}}}]{Hojjati_Pogosian}
\bibinfo{author}{\bibfnamefont{A.}~\bibnamefont{{Hojjati}}},
  \bibinfo{author}{\bibfnamefont{L.}~\bibnamefont{{Pogosian}}},
  \bibnamefont{and} \bibinfo{author}{\bibfnamefont{G.-B.}
  \bibnamefont{{Zhao}}}, \bibinfo{journal}{ArXiv e-prints}
  (\bibinfo{year}{2011}{\natexlab{b}}), \eprint{1106.4543}.

\bibitem[{\citenamefont{{Dossett} et~al.}(2011)\citenamefont{{Dossett},
  {Ishak}, and {Moldenhauer}}}]{Dossett_2011}
\bibinfo{author}{\bibfnamefont{J.}~\bibnamefont{{Dossett}}},
  \bibinfo{author}{\bibfnamefont{M.}~\bibnamefont{{Ishak}}}, \bibnamefont{and}
  \bibinfo{author}{\bibfnamefont{J.}~\bibnamefont{{Moldenhauer}}},
  \bibinfo{journal}{ArXiv e-prints}  (\bibinfo{year}{2011}),
  \eprint{1109.4583}.

\bibitem[{\citenamefont{{Sotiriou} and {Faraoni}}(2010)}]{Sotiriou_fR}
\bibinfo{author}{\bibfnamefont{T.~P.} \bibnamefont{{Sotiriou}}}
  \bibnamefont{and}
  \bibinfo{author}{\bibfnamefont{V.}~\bibnamefont{{Faraoni}}},
  \bibinfo{journal}{Reviews of Modern Physics} \textbf{\bibinfo{volume}{82}},
  \bibinfo{pages}{451} (\bibinfo{year}{2010}), \eprint{0805.1726}.

\bibitem[{\citenamefont{{Larson} et~al.}(2011)}]{wmap7}
\bibinfo{author}{\bibfnamefont{D.}~\bibnamefont{{Larson}}}
  \bibnamefont{et~al.}, \bibinfo{journal}{ApJS} \textbf{\bibinfo{volume}{192}},
  \bibinfo{pages}{16} (\bibinfo{year}{2011}), \eprint{arXiv:1001.4635}.

\bibitem[{\citenamefont{{Reid} et~al.}(2010)}]{sdss-dr7}
\bibinfo{author}{\bibfnamefont{B.~A.} \bibnamefont{{Reid}}}
  \bibnamefont{et~al.}, \bibinfo{journal}{MNRAS}
  \textbf{\bibinfo{volume}{404}}, \bibinfo{pages}{60} (\bibinfo{year}{2010}),
  \eprint{arXiv:0907.1659}.

\bibitem[{\citenamefont{{Riess} et~al.}(2011)}]{riess2011}
\bibinfo{author}{\bibfnamefont{A.~G.} \bibnamefont{{Riess}}}
  \bibnamefont{et~al.}, \bibinfo{journal}{\apj} \textbf{\bibinfo{volume}{730}},
  \bibinfo{pages}{119} (\bibinfo{year}{2011}), \eprint{arXiv:1103.2976}.

\bibitem[{\citenamefont{{Hamann} et~al.}(2008)\citenamefont{{Hamann},
  {Lesgourgues}, and {Mangano}}}]{bbn}
\bibinfo{author}{\bibfnamefont{J.}~\bibnamefont{{Hamann}}},
  \bibinfo{author}{\bibfnamefont{J.}~\bibnamefont{{Lesgourgues}}},
  \bibnamefont{and}
  \bibinfo{author}{\bibfnamefont{G.}~\bibnamefont{{Mangano}}},
  \bibinfo{journal}{JCAP} \textbf{\bibinfo{volume}{3}}, \bibinfo{pages}{4}
  (\bibinfo{year}{2008}), \eprint{arXiv:0712.2826}.

\bibitem[{\citenamefont{{Amanullah} et~al.}(2010)}]{Amanullah2010}
\bibinfo{author}{\bibfnamefont{R.}~\bibnamefont{{Amanullah}}}
  \bibnamefont{et~al.}, \bibinfo{journal}{\apj} \textbf{\bibinfo{volume}{716}},
  \bibinfo{pages}{712} (\bibinfo{year}{2010}), \eprint{arXiv:1004.1711}.

\bibitem[{\citenamefont{{Sachs} and {Wolfe}}(1967)}]{Sachs_Wolfe}
\bibinfo{author}{\bibfnamefont{R.~K.} \bibnamefont{{Sachs}}} \bibnamefont{and}
  \bibinfo{author}{\bibfnamefont{A.~M.} \bibnamefont{{Wolfe}}},
  \bibinfo{journal}{\apj} \textbf{\bibinfo{volume}{147}}, \bibinfo{pages}{73}
  (\bibinfo{year}{1967}).

\bibitem[{\citenamefont{Crittenden and Turok}(1996)}]{Crittenden_Turok}
\bibinfo{author}{\bibfnamefont{R.~G.} \bibnamefont{Crittenden}}
  \bibnamefont{and} \bibinfo{author}{\bibfnamefont{N.}~\bibnamefont{Turok}},
  \bibinfo{journal}{Phys. Rev. Lett.} \textbf{\bibinfo{volume}{76}},
  \bibinfo{pages}{575} (\bibinfo{year}{1996}),
  \urlprefix\url{http://link.aps.org/doi/10.1103/PhysRevLett.76.575}.

\bibitem[{\citenamefont{Zhang}(2006)}]{Zhang_2006}
\bibinfo{author}{\bibfnamefont{P.}~\bibnamefont{Zhang}},
  \bibinfo{journal}{Phys. Rev. D} \textbf{\bibinfo{volume}{73}},
  \bibinfo{pages}{123504} (\bibinfo{year}{2006}),
  \urlprefix\url{http://link.aps.org/doi/10.1103/PhysRevD.73.123504}.

\bibitem[{\citenamefont{{Giannantonio}
  et~al.}(2010)\citenamefont{{Giannantonio}, {Martinelli}, {Silvestri}, and
  {Melchiorri}}}]{Giannantonio_2010}
\bibinfo{author}{\bibfnamefont{T.}~\bibnamefont{{Giannantonio}}},
  \bibinfo{author}{\bibfnamefont{M.}~\bibnamefont{{Martinelli}}},
  \bibinfo{author}{\bibfnamefont{A.}~\bibnamefont{{Silvestri}}},
  \bibnamefont{and}
  \bibinfo{author}{\bibfnamefont{A.}~\bibnamefont{{Melchiorri}}},
  \bibinfo{journal}{JCAP} \textbf{\bibinfo{volume}{4}}, \bibinfo{pages}{30}
  (\bibinfo{year}{2010}), \eprint{0909.2045}.

\bibitem[{\citenamefont{Dodelson}(2003)}]{Dodelson}
\bibinfo{author}{\bibfnamefont{S.}~\bibnamefont{Dodelson}},
  \emph{\bibinfo{title}{Modern Cosmology}} (\bibinfo{publisher}{Academic
  Press}, \bibinfo{year}{2003}).

\bibitem[{\citenamefont{Brans and Dicke}(1961)}]{BransDicke}
\bibinfo{author}{\bibfnamefont{C.}~\bibnamefont{Brans}} \bibnamefont{and}
  \bibinfo{author}{\bibfnamefont{R.~H.} \bibnamefont{Dicke}},
  \bibinfo{journal}{Phys. Rev.} \textbf{\bibinfo{volume}{124}},
  \bibinfo{pages}{925} (\bibinfo{year}{1961}).

\bibitem[{\citenamefont{{Koivisto} and {Mota}}(2006)}]{Koivisto_Mota}
\bibinfo{author}{\bibfnamefont{T.}~\bibnamefont{{Koivisto}}} \bibnamefont{and}
  \bibinfo{author}{\bibfnamefont{D.~F.} \bibnamefont{{Mota}}},
  \bibinfo{journal}{\prd} \textbf{\bibinfo{volume}{73}},
  \bibinfo{pages}{083502} (\bibinfo{year}{2006}),
  \eprint{arXiv:astro-ph/0512135}.

\bibitem[{\citenamefont{{Mota} et~al.}(2007)\citenamefont{{Mota},
  {Kristiansen}, {Koivisto}, and {Groeneboom}}}]{Mota_2007}
\bibinfo{author}{\bibfnamefont{D.~F.} \bibnamefont{{Mota}}},
  \bibinfo{author}{\bibfnamefont{J.~R.} \bibnamefont{{Kristiansen}}},
  \bibinfo{author}{\bibfnamefont{T.}~\bibnamefont{{Koivisto}}},
  \bibnamefont{and} \bibinfo{author}{\bibfnamefont{N.~E.}
  \bibnamefont{{Groeneboom}}}, \bibinfo{journal}{MNRAS}
  \textbf{\bibinfo{volume}{382}}, \bibinfo{pages}{793} (\bibinfo{year}{2007}),
  \eprint{0708.0830}.

\bibitem[{\citenamefont{Horava}(2009{\natexlab{a}})}]{HL1}
\bibinfo{author}{\bibfnamefont{P.}~\bibnamefont{Horava}},
  \bibinfo{journal}{Phys. Rev. Lett.} \textbf{\bibinfo{volume}{102}},
  \bibinfo{pages}{161301} (\bibinfo{year}{2009}{\natexlab{a}}).

\bibitem[{\citenamefont{Horava}(2009{\natexlab{b}})}]{HL2}
\bibinfo{author}{\bibfnamefont{P.}~\bibnamefont{Horava}},
  \bibinfo{journal}{Phys. Rev. D} \textbf{\bibinfo{volume}{79}},
  \bibinfo{pages}{084008} (\bibinfo{year}{2009}{\natexlab{b}}).

\bibitem[{\citenamefont{{Zlosnik} et~al.}(2008)\citenamefont{{Zlosnik},
  {Ferreira}, and {Starkman}}}]{Zlosnik_structure_growth}
\bibinfo{author}{\bibfnamefont{T.~G.} \bibnamefont{{Zlosnik}}},
  \bibinfo{author}{\bibfnamefont{P.~G.} \bibnamefont{{Ferreira}}},
  \bibnamefont{and} \bibinfo{author}{\bibfnamefont{G.~D.}
  \bibnamefont{{Starkman}}}, \bibinfo{journal}{\prd}
  \textbf{\bibinfo{volume}{77}}, \bibinfo{pages}{084010}
  (\bibinfo{year}{2008}), \eprint{0711.0520}.

\bibitem[{\citenamefont{Ba\~nados et~al.}(2009)\citenamefont{Ba\~nados,
  Ferreira, and Skordis}}]{Banados_Ferreira_Skordis}
\bibinfo{author}{\bibfnamefont{M.}~\bibnamefont{Ba\~nados}},
  \bibinfo{author}{\bibfnamefont{P.~G.} \bibnamefont{Ferreira}},
  \bibnamefont{and} \bibinfo{author}{\bibfnamefont{C.}~\bibnamefont{Skordis}},
  \bibinfo{journal}{Phys. Rev. D} \textbf{\bibinfo{volume}{79}},
  \bibinfo{pages}{063511} (\bibinfo{year}{2009}).

\bibitem[{\citenamefont{Skordis}(2009)}]{Skordis}
\bibinfo{author}{\bibfnamefont{C.}~\bibnamefont{Skordis}},
  \bibinfo{journal}{Phys. Rev. D} \textbf{\bibinfo{volume}{79}},
  \bibinfo{pages}{123527} (\bibinfo{year}{2009}).

\bibitem[{\citenamefont{{Wands} et~al.}(2000)\citenamefont{{Wands}, {Malik},
  {Lyth}, and {Liddle}}}]{Wands_etal_2000}
\bibinfo{author}{\bibfnamefont{D.}~\bibnamefont{{Wands}}},
  \bibinfo{author}{\bibfnamefont{K.~A.} \bibnamefont{{Malik}}},
  \bibinfo{author}{\bibfnamefont{D.~H.} \bibnamefont{{Lyth}}},
  \bibnamefont{and} \bibinfo{author}{\bibfnamefont{A.~R.}
  \bibnamefont{{Liddle}}}, \bibinfo{journal}{\prd}
  \textbf{\bibinfo{volume}{62}}, \bibinfo{pages}{043527}
  (\bibinfo{year}{2000}), \eprint{arXiv:astro-ph/0003278}.

\bibitem[{\citenamefont{{Cardoso} and {Wands}}(2008)}]{Cardoso_Wands}
\bibinfo{author}{\bibfnamefont{A.}~\bibnamefont{{Cardoso}}} \bibnamefont{and}
  \bibinfo{author}{\bibfnamefont{D.}~\bibnamefont{{Wands}}},
  \bibinfo{journal}{\prd} \textbf{\bibinfo{volume}{77}},
  \bibinfo{pages}{123538} (\bibinfo{year}{2008}), \eprint{0801.1667}.

\bibitem[{\citenamefont{{Bertschinger}}(2006)}]{Bertschinger2006}
\bibinfo{author}{\bibfnamefont{E.}~\bibnamefont{{Bertschinger}}},
  \bibinfo{journal}{\apj} \textbf{\bibinfo{volume}{648}}, \bibinfo{pages}{797}
  (\bibinfo{year}{2006}), \eprint{arXiv:astro-ph/0604485}.

\bibitem[{\citenamefont{{Komatsu} et~al.}(2011)\citenamefont{{Komatsu},
  {Smith}, {Dunkley}, {Bennett}, {Gold}, {Hinshaw}, {Jarosik}, {Larson},
  {Nolta}, {Page} et~al.}}]{WMAP7_Komatsu}
\bibinfo{author}{\bibfnamefont{E.}~\bibnamefont{{Komatsu}}},
  \bibinfo{author}{\bibfnamefont{K.~M.} \bibnamefont{{Smith}}},
  \bibinfo{author}{\bibfnamefont{J.}~\bibnamefont{{Dunkley}}},
  \bibinfo{author}{\bibfnamefont{C.~L.} \bibnamefont{{Bennett}}},
  \bibinfo{author}{\bibfnamefont{B.}~\bibnamefont{{Gold}}},
  \bibinfo{author}{\bibfnamefont{G.}~\bibnamefont{{Hinshaw}}},
  \bibinfo{author}{\bibfnamefont{N.}~\bibnamefont{{Jarosik}}},
  \bibinfo{author}{\bibfnamefont{D.}~\bibnamefont{{Larson}}},
  \bibinfo{author}{\bibfnamefont{M.~R.} \bibnamefont{{Nolta}}},
  \bibinfo{author}{\bibfnamefont{L.}~\bibnamefont{{Page}}},
  \bibnamefont{et~al.}, \bibinfo{journal}{ApJS} \textbf{\bibinfo{volume}{192}},
  \bibinfo{pages}{18} (\bibinfo{year}{2011}), \eprint{1001.4538}.

\bibitem[{\citenamefont{{Nesseris} and
  {Perivolaropoulos}}(2007)}]{Nesseris_Perivo}
\bibinfo{author}{\bibfnamefont{S.}~\bibnamefont{{Nesseris}}} \bibnamefont{and}
  \bibinfo{author}{\bibfnamefont{L.}~\bibnamefont{{Perivolaropoulos}}},
  \bibinfo{journal}{JCAP} \textbf{\bibinfo{volume}{1}}, \bibinfo{pages}{18}
  (\bibinfo{year}{2007}), \eprint{arXiv:astro-ph/0610092}.

\bibitem[{\citenamefont{Perivolaropoulos}(2005)}]{Perivolaropoulos}
\bibinfo{author}{\bibfnamefont{L.}~\bibnamefont{Perivolaropoulos}},
  \bibinfo{journal}{Phys. Rev. D} \textbf{\bibinfo{volume}{71}},
  \bibinfo{pages}{063503} (\bibinfo{year}{2005}),
  \urlprefix\url{http://link.aps.org/doi/10.1103/PhysRevD.71.063503}.

\bibitem[{\citenamefont{{Cline} et~al.}(2004)\citenamefont{{Cline}, {Jeon}, and
  {Moore}}}]{Cline_2004}
\bibinfo{author}{\bibfnamefont{J.~M.} \bibnamefont{{Cline}}},
  \bibinfo{author}{\bibfnamefont{S.}~\bibnamefont{{Jeon}}}, \bibnamefont{and}
  \bibinfo{author}{\bibfnamefont{G.~D.} \bibnamefont{{Moore}}},
  \bibinfo{journal}{\prd} \textbf{\bibinfo{volume}{70}},
  \bibinfo{pages}{043543} (\bibinfo{year}{2004}),
  \eprint{arXiv:hep-ph/0311312}.

\bibitem[{\citenamefont{{Gannouji} et~al.}(2006)\citenamefont{{Gannouji},
  {Polarski}, {Ranquet}, and {Starobinsky}}}]{Gannouji_Polarski}
\bibinfo{author}{\bibfnamefont{R.}~\bibnamefont{{Gannouji}}},
  \bibinfo{author}{\bibfnamefont{D.}~\bibnamefont{{Polarski}}},
  \bibinfo{author}{\bibfnamefont{A.}~\bibnamefont{{Ranquet}}},
  \bibnamefont{and} \bibinfo{author}{\bibfnamefont{A.~A.}
  \bibnamefont{{Starobinsky}}}, \bibinfo{journal}{JCAP}
  \textbf{\bibinfo{volume}{9}}, \bibinfo{pages}{16} (\bibinfo{year}{2006}),
  \eprint{arXiv:astro-ph/0606287}.

\bibitem[{\citenamefont{{Abdalla} et~al.}(2005)\citenamefont{{Abdalla},
  {Nojiri}, and {Odintsov}}}]{Abdalla_Nojiri_Odintsov}
\bibinfo{author}{\bibfnamefont{M.~C.~B.} \bibnamefont{{Abdalla}}},
  \bibinfo{author}{\bibfnamefont{S.}~\bibnamefont{{Nojiri}}}, \bibnamefont{and}
  \bibinfo{author}{\bibfnamefont{S.~D.} \bibnamefont{{Odintsov}}},
  \bibinfo{journal}{Classical and Quantum Gravity}
  \textbf{\bibinfo{volume}{22}}, \bibinfo{pages}{L35} (\bibinfo{year}{2005}),
  \eprint{hep-th/0409177}.

\bibitem[{\citenamefont{{Amendola} and {Tsujikawa}}(2008)}]{Amendola_Tsujikawa}
\bibinfo{author}{\bibfnamefont{L.}~\bibnamefont{{Amendola}}} \bibnamefont{and}
  \bibinfo{author}{\bibfnamefont{S.}~\bibnamefont{{Tsujikawa}}},
  \bibinfo{journal}{Physics Letters B} \textbf{\bibinfo{volume}{660}},
  \bibinfo{pages}{125} (\bibinfo{year}{2008}), \eprint{0705.0396}.

\bibitem[{\citenamefont{{Martin} et~al.}(2006)\citenamefont{{Martin}, {Schimd},
  and {Uzan}}}]{Martin_Schimd_Uzan}
\bibinfo{author}{\bibfnamefont{J.}~\bibnamefont{{Martin}}},
  \bibinfo{author}{\bibfnamefont{C.}~\bibnamefont{{Schimd}}}, \bibnamefont{and}
  \bibinfo{author}{\bibfnamefont{J.-P.} \bibnamefont{{Uzan}}},
  \bibinfo{journal}{Physical Review Letters} \textbf{\bibinfo{volume}{96}},
  \bibinfo{pages}{061303} (\bibinfo{year}{2006}),
  \eprint{arXiv:astro-ph/0510208}.

\bibitem[{\citenamefont{{Libanov} et~al.}(2007)\citenamefont{{Libanov},
  {Rubakov}, {Papantonopoulos}, {Sami}, and {Tsujikawa}}}]{Libanov_Rubakov}
\bibinfo{author}{\bibfnamefont{M.}~\bibnamefont{{Libanov}}},
  \bibinfo{author}{\bibfnamefont{V.}~\bibnamefont{{Rubakov}}},
  \bibinfo{author}{\bibfnamefont{E.}~\bibnamefont{{Papantonopoulos}}},
  \bibinfo{author}{\bibfnamefont{M.}~\bibnamefont{{Sami}}}, \bibnamefont{and}
  \bibinfo{author}{\bibfnamefont{S.}~\bibnamefont{{Tsujikawa}}},
  \bibinfo{journal}{JCAP} \textbf{\bibinfo{volume}{8}}, \bibinfo{pages}{10}
  (\bibinfo{year}{2007}), \eprint{0704.1848}.

\bibitem[{\citenamefont{{Kunz} and {Sapone}}(2006)}]{Kunz_Sapone_phantom}
\bibinfo{author}{\bibfnamefont{M.}~\bibnamefont{{Kunz}}} \bibnamefont{and}
  \bibinfo{author}{\bibfnamefont{D.}~\bibnamefont{{Sapone}}},
  \bibinfo{journal}{\prd} \textbf{\bibinfo{volume}{74}},
  \bibinfo{pages}{123503} (\bibinfo{year}{2006}),
  \eprint{arXiv:astro-ph/0609040}.

\bibitem[{\citenamefont{Zhao et~al.}(2010)\citenamefont{Zhao, Giannantonio,
  Pogosian, Silvestri, Bacon et~al.}}]{Zhao2010}
\bibinfo{author}{\bibfnamefont{G.-B.} \bibnamefont{Zhao}},
  \bibinfo{author}{\bibfnamefont{T.}~\bibnamefont{Giannantonio}},
  \bibinfo{author}{\bibfnamefont{L.}~\bibnamefont{Pogosian}},
  \bibinfo{author}{\bibfnamefont{A.}~\bibnamefont{Silvestri}},
  \bibinfo{author}{\bibfnamefont{D.~J.} \bibnamefont{Bacon}},
  \bibnamefont{et~al.}, \bibinfo{journal}{Phys.Rev.}
  \textbf{\bibinfo{volume}{D81}}, \bibinfo{pages}{103510}
  (\bibinfo{year}{2010}), \eprint{1003.0001}.

\bibitem[{\citenamefont{{Tsujikawa}}(2007)}]{Tsujikawa_2007}
\bibinfo{author}{\bibfnamefont{S.}~\bibnamefont{{Tsujikawa}}},
  \bibinfo{journal}{\prd} \textbf{\bibinfo{volume}{76}},
  \bibinfo{pages}{023514} (\bibinfo{year}{2007}), \eprint{0705.1032}.

\bibitem[{\citenamefont{{de Felice} et~al.}(2010)\citenamefont{{de Felice},
  {Mukohyama}, and {Tsujikawa}}}]{deFelice_2010}
\bibinfo{author}{\bibfnamefont{A.}~\bibnamefont{{de Felice}}},
  \bibinfo{author}{\bibfnamefont{S.}~\bibnamefont{{Mukohyama}}},
  \bibnamefont{and}
  \bibinfo{author}{\bibfnamefont{S.}~\bibnamefont{{Tsujikawa}}},
  \bibinfo{journal}{\prd} \textbf{\bibinfo{volume}{82}},
  \bibinfo{pages}{023524} (\bibinfo{year}{2010}), \eprint{1006.0281}.

\end{thebibliography}

\end{document}